\magnification\magstephalf
%\magnification\magstep1
\def\Im{\,\hbox{Im}\,}
\overfullrule 0pt
\input epsf

\font\rfont=cmr9 at 9 true pt
\def\ref#1{$^{\hbox{\rfont {[#1]}}}$}
\def\R{{\hbox{{\sevenrm R}}}}

\def\eins{\hbox{$1 \hskip-1.0mm  {\rm l}$}}

\def\tr{\rm tr}

  %%Fonts

\font\fourteenbf=cmbx12 scaled\magstep1

\font\tenbfit=cmbxti10
\font\sevenbfit=cmbxti10 at 7pt
\font\fivebfit=cmbxti10 at 5pt
\newfam\bfitfam \def\bfit{\fam\bfitfam\tenbfit}
\textfont\bfitfam=\tenbfit  \scriptfont\bfitfam=\sevenbfit
\scriptscriptfont\bfitfam=\fivebfit

\font\eightrm=cmr8

\font\eightit=cmti8

  %%Greek
 \def\b {\beta}  
\def\e{\epsilon}  \def\o{\omega}
  
\def\T {\Theta}  \def\G {\Gamma}
\def\pd {\partial}
\def\pmb#1{\setbox0=\hbox{#1}% \kern-.025em\copy0\kern-\wd0
 \kern.05em\copy0\kern-\wd0 \kern-.025em\raise.0433em\box0 }

\def\btau{{\pmb{$ \tau$}}}
\def\bchi{{\pmb{$ \chi$}}}
\def\slash{/\kern-.5em}
\def\2{\prime 2}
\def\Rpi{\pi}

  %%Fractions
\def \half {{\textstyle {1 \over 2}}}

 %

 %%FORMATTING

\def\boxit#1{\vbox{\hrule\hbox{\vrule\kern1pt\vbox
{\kern1pt#1\kern1pt}\kern1pt\vrule}\hrule}}

\def\h{\hfill\break}
\parskip=6pt
\parindent=0pt
\hsize=17truecm\hoffset=-5truemm
\vsize=23truecm
\def\footnoterule{\kern-3pt
\hrule width 17truecm \kern 2.6pt}

  %%REFERENCES
%     \defref\label{text}
% generates a number, assigns it to \label, generates an entry.
% To list the refs,  \listrefs
% (Extracted and adapted from harvmac.tex by P Ginsparg)

\catcode`\@=11 % This allows us to modify PLAIN macros.

\def\nolabels{\def\wrlabeL##1{}\def\eqlabeL##1{}\def\reflabeL##1{}}
\def\writelabels{\def\wrlabeL##1{\leavevmode\vadjust{\rlap{\smash%
{\line{{\escapechar=` \hfill\rlap{\sevenrm\hskip.03in\string##1}}}}}}}%
\def\eqlabeL##1{{\escapechar-1\rlap{\sevenrm\hskip.05in\string##1}}}%
\def\reflabeL##1{\noexpand\llap{\noexpand\sevenrm\string\string\string##1}}}
\nolabels
\global\newcount\refno \global\refno=1
\newwrite\rfile
\def\defref{$^{{\hbox{\rfont [\the\refno]}}}$\nref}
\def\nref#1{\xdef#1{\the\refno}\writedef{#1\leftbracket#1}%
\ifnum\refno=1\immediate\openout\rfile=refs.tmp\fi
\global\advance\refno by1\chardef\wfile=\rfile\immediate
\write\rfile{\noexpand\item{#1\ }\reflabeL{#1\hskip.31in}\pctsign}\findarg}
%       horrible hack to sidestep tex \write limitation
\def\findarg#1#{\begingroup\obeylines\newlinechar=`\^^M\pass@rg}
{\obeylines\gdef\pass@rg#1{\writ@line\relax #1^^M\hbox{}^^M}%
\gdef\writ@line#1^^M{\expandafter\toks0\expandafter{\striprel@x #1}%
\edef\next{\the\toks0}\ifx\next\em@rk\let\next=\endgroup\else\ifx\next\empty%
\else\immediate\write\wfile{\the\toks0}\fi\let\next=\writ@line\fi\next\relax}}
\def\striprel@x#1{} \def\em@rk{\hbox{}}
\def\lref{\begingroup\obeylines\lr@f}
\def\lr@f#1#2{\gdef#1{\defref#1{#2}}\endgroup\unskip}
\newwrite\lfile
{\escapechar-1\xdef\pctsign{\string\%}\xdef\leftbracket{\string\{}
\xdef\rightbracket{\string\}}}

\def\writestop{\def\writestoppt{\immediate\write\lfile{\string\p
ageno%
\the\pageno\string\startrefs\leftbracket\the\refno\rightbracket%
\string\def\string\secsym\leftbracket\secsym\rightbracket%
\string\secno\the\secno\string\meqno\the\meqno}\immediate\closeout\lfile}}
\def\writestoppt{}\def\writedef#1{}
\catcode`\@=12 % at signs are no longer letters

\rightline{{}\hfill
DAMTP 98/72}
\rightline{HD-THEP-98-26}
\rightline{TUW 98-15}
\vskip 10mm

{\fourteenbf
\centerline{RENORMALISATION OF THE}
\vskip 2mm
\centerline{ NONPERTURBATIVE THERMAL
PRESSURE}}
\bigskip

\centerline{D B\"odeker$^1$, P V Landshoff$^2$, O Nachtmann$^1$
and A Rebhan$^3$}
\smallskip
\centerline{\it Institut f\"ur Theoretische Physik,
Universit\"at Heidelberg$^1$}
\centerline{\it DAMTP, University of Cambridge$^2$}
\centerline{\it Institut f\"ur Theoretische Physik, Technische Universit\"at
Wien$^3$}
\vskip 2mm
{\eightrm
\centerline{bodeker@thphys.uni-heidelberg.de \ \
p.v.landshoff@damtp.cambridge.ac.uk}
\centerline{o.nachtmann@thphys.uni-heidelberg.de\ \
rebhana@tph16.tuwien.ac.at}}

\vskip 10mm
\midinsert\leftskip 8mm\rightskip 8mm
{\bf Abstract}

We show how the fully
resummed thermal pressure is rendered ultraviolet finite
by standard zero-temperature renormalisation. The analysis is developed in
a 6-dimensional scalar
model that mimics QED and has $N$ flavours. The $N\to\infty$ limit of the
model can be calculated completely. 
At a critical temperature, one of
the degrees of freedom has vanishing screening mass like
the transverse gauge bosons in four-dimensional finite-temperature 
perturbation
theory.
The renormalised nonperturbative interaction pressure of this
model is evaluated numerically.

\endinsert
\vskip 8mm
{\bf 1 Introduction}

The perturbation series for the pressure in finite-temperature QCD suffers
from severe infrared problems. In principle, these may be cured
by a resummation technique. This resummation is most simply carried
out\defref\first{
I T Drummond, R R Horgan, P V Landshoff and A Rebhan, Physics Letters
B398 (1997) 326
}
before renormalisation. It goes without saying that, after renormalisation,
the pressure must be finite if it is to make physical sense, but it is
far from obvious how the mathematics takes care of this. In this paper we
study this in a model that may be regarded as a simplified mimic of QED,
in which there are $N$  particles each having the same mass and ``charge''.
Our analysis makes use of standard techniques for renormalising
composite operators. We find that indeed the usual renormalisation, carried
out purely at the zero-temperature level and therefore introducing no new
quantities needing to be determined by experiment, renders the pressure
finite.

For simplicity, our discussion begins with
the $N\to\infty$ limit of the model, which can be calculated exactly.
The model is richer than the
large-$N$ $\phi ^4$ theory which we have studied previously\defref\second{
I T Drummond, R R Horgan, P V Landshoff and A Rebhan, hep-ph/9708426,
Nucl Phys B524 (1998)
}, in that now the self-energies in the large-$N$ limit vary with momentum,
and wave-function renormalisation is needed. After renormalising
the expression for the pressure and going some way towards evaluating
it analytically, we complete the calculation of the large-$N$ case
numerically.

Such a calculation is potentially useful even when $N$ is not large.
The formula for the resummed pressure involves the thermal self-energies
of the fields, which inevitably are calculated in some approximation from
a finite number of Feynman graphs.
The resummation then effectively converts this finite set of graphs to a
contribution to the pressure from an infinite number of graphs.
While the exact form of the pressure must
be ultraviolet finite, it is not obvious that it is still finite when
only a partial set of graphs is included in the self-energy. One way of
selecting a consistent approximation is to use in the finite-$N$ case
only the set of graphs that would survive to some given order in $N^{-1}$
if one were to take the large-$N$ limit.
\def\T{{\rm T}}
In a theory in which there are a number of real
unrenormalised fields $\phi _r$,
the pressure at temperature $T$ is calculated\ref{\first} from the thermal
averages of the composite operators $\phi _r^2$:
$$
{\partial\over\partial m^2_{0r}}P(T)=
- \half \Big (\big\langle \phi _r(x)\phi _r(0)\big\rangle _T
-\langle 0|\phi _r(x) \phi _r(0)|0\rangle\Big )_{x=0}
\eqno(1.1a)
$$
or, equivalently,
$$
{\partial\over\partial m^2_{0r}}P(T)=
- \half \Big (\big\langle\T\phi _r(x) \phi _r(0)\big\rangle _T
-\langle 0|\T\phi _r(x)\phi _r(0)|0\rangle\Big )_{x=0}
\eqno(1.1b)
$$
Here, the differentiation is with respect to the unrenormalised mass of
the field $\phi _r$, with all the other unrenormalised parameters kept fixed.
One way to integrate this to give $P(T)$ is to write
$$
m^{\2}_{0r}=xm^2_{0r}~~~~~~~~~~~~~r=1,2,\dots$$$$
{\hbox{d}\over\hbox{d} x}P(T)=
\sum _r m^2_{0r}{\partial\over\partial m^{'2}_{0r}}P(T)
\eqno(1.2)
$$
Then integrate with respect to $x$ from 1 to $\infty$ and insert the
boundary condition that the pressure should vanish when $x=\infty$,
that is when all the masses are infinite\footnote{$^*$}{
Our methods resemble those of a renormalisation-flow
analysis\defref\wet{
C Wetterich, Physics Letters  B301 (1993) 90
}.}. However if, as is the case in
QED, it happens that taking just one of the masses  --- $m_{0s}$ say ---
to infinity switches off all the interaction, there is a simpler method:
$$
P(T)=-\int _{m_{0s}^2}^{\infty}dm^{\2}_{0s}
{\partial\over\partial m^{\2}_{0s}}P(T)
+P_0(T)
\eqno(1.3)
$$
where $P_0(T)$ is the contribution to the thermal
pressure from all the fields except
$\phi _s$, with the interaction between them switched off.

The thermal averages of the composite operators that appear
in (1.1) may be expressed as integrals over thermal Green's functions:
$$
\big\langle \phi _r(x)\phi _r(0)\big\rangle _T\Big\arrowvert _{x=0}=
\int {d^nq\over (2\pi )^n}D_{rT}^{12}(q)
$$$$
\big\langle\T\phi _r(x)\phi _r(0)\big\rangle _T\Big\arrowvert _{x=0}=
\int {d^nq\over (2\pi )^n}D_{rT}^{11}(q)
\eqno (1.4)
$$
The notation $D_{rT}^{12}$, $D_{rT}^{11}$ is that of real-time thermal field 
theory in
the Keldysh formalism\defref\lebellac{
M LeBellac, {\it Thermal field theory}, Cambridge University Press (1997)
} with a time path with $\sigma = 0$: 
they are elements of the matrix propagator
$$
{\bf D}_{rT}(q)={\bf M}_T(q^0)\left [
\matrix{D_{rT}(q)& 0\cr
0& D^*_{rT}(q)\cr}\right ] {\bf M}_T(q^0)
\eqno(1.5a)
$$
where
$$
D_{rT}(q)={i\over q^2-m_{0r}^2-\Pi _{rT}(q^0,{\bf q}^2)}$$$$
{\bf M}_T(q^0)=\left[{1\over e^{|q^0|/T}-1}\right]^{1/2}
\left [\matrix{e^{\half |q^0|/T}&e^{-\half q^0/T}\cr
                              e^{\half q^0/T}&e^{\half |q^0|/T}\cr}\right ]
\eqno(1.5b)
$$
Here,  $\Pi _{rT}$ is defined in terms of the thermal self-energy
matrix
$$
-i{\bf\Pi}_{rT}={\bf M}^{-1}_T(q^0)\left [
\matrix{-i\Pi _{rT}& 0\cr
0& (-i\Pi_{rT}(q))^*\cr}\right ] {\bf M}^{-1}_T(q^0)
\eqno(1.5c)
$$
Inserting (1.5) into (1.1a) gives
$$
{\partial\over\partial m^2_{0r}}P(T)=\half\hbox{ Im }
\int {d^nq\over (2\pi )^n}
\Big\{{1+2n(q)\over q^2-m_{0r}^2-\Pi _{rT} (q^0,{\bf q}^2)}-{1\over
q^2-m_{0r}^2-\Pi _r (q^2)}
\Big\}
\eqno(1.1c)
$$
where $\Pi _r (q^2)$ is the zero-temperature self energy and $n(q)$ is the
Bose distribution $(e^{|q^0|/T}-1)^{-1}$. The version
(1.1b) of our basic formula gives instead
$${\partial\over\partial m^2_{0r}}P(T)=\half\int{d^nq\over
(2\pi)^n}\Big\{\hbox{ Im }{2n(q)\over q^2-m^2_{0r}-\Pi_{rT}(q^0,{\bf q}^2)}
~~~~~~~~~~~~~~~~~~~~~~~~~~~~~~~~~~~~~~~~~~$$$$
~~~~~~~~~~~~~~~~~~~~~~~~~~~~~~~~~~~~
+{i\over q^2-m^2_{0r}-\Pi_r(q^2)}-{i\over q^2-m^2_{0r}-\Pi_{rT}(q^0,{\bf q}^2)}
\Big\}\eqno(1.1d)$$
The equivalence of (1.1c) and (1.1d) may be seen from the fact\footnote
{*}{Formula (2.68) of reference [3]} that each of the last
two terms in the integrand of (1.1d) is the analytic continuation
in $q^0$ of the real-time propagator, and by making a Wick
rotation, so verifying that the integral over each of the last
two terms is real.

We apply the formula (1.1) to a mock electron-photon interaction
in which, for simplicity, the fields are scalar; its unrenormalised form is
$$
{\cal L}^{\hbox{{\sevenrm INT}}}=-\lambda _0
A^a\psi _{r} ^{\dag}\tau ^a\psi _{r}
\eqno(1.6)
$$
The masses are
$m_{01}$ for the electrons, and $m_{02}$ for the photon.
In proper QED, $C$-parity or spin conservation removes one-photon-reducible
graphs from the pressure; here we achieve this instead by making the
photon an isovector, and the electrons isodoublets, so that
$a$ runs over 3 values with $\tau^a$ the Pauli matrices.
We take space-time to be 6-dimensional, so
that $\lambda _0$ is dimensionless
and the divergences are similar to those of proper QED.
In intermediate steps we use dimensional regularisation
with $n=6-2\epsilon$.

There are $N$ identical electrons: the index
$r$ runs over $N$ values. In the next section we set
$$
\lambda _0={g_0\over\surd N}
\eqno(1.7)
$$
and consider the case of large $N$.
The more general case, where $N$ is not necessarily large, is the subject
of section 3.
In section 4 we return to the large-N version of the model
and evaluate the leading term in the interaction pressure completely.
The thermal ``photon'' spectrum turns out to involve negative corrections
to the mass such that there is a critical temperature, where
screening disappears while keeping the plasmon mass nonzero. 
Right at the critical temperature,
where the nonperturbative interaction pressure is still well-defined,
the spectrum of our model is even rather similar to that of
perturbative four-dimensional
gauge theories in that it has a vanishing screening mass
like the magnetostatic modes.
Section 5 is a summary and discussion.
\bigskip
\goodbreak
{\bf 2 Large} {\bfit N}

In the large-$N$ limit, the
free-field pressure is linear in $N$, and the correction from the
interaction is of order $N^0$. To calculate this we need the leading terms
in the photon self-energy $g_0^2\pi\delta ^{ab}$
and the electron self-energy, which
are respectively of order $N^0$ and $N^{-1}$ and correspond to the graphs
of figure 1. In figure 1b, the photon propagator is the Dyson-resummed
propagator with the photon self energy $g_0^2\pi$
of figure 1a.

\topinsert
\centerline{{\epsfxsize=60mm\epsfbox{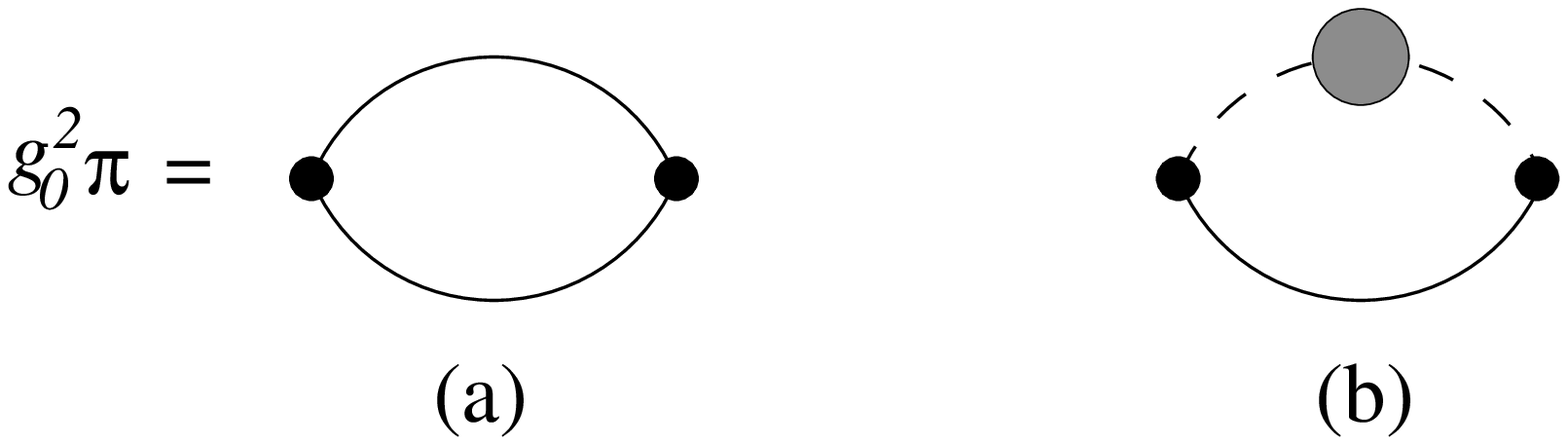}}}
\vskip 1pt
\centerline{Figure 1: self-energy graphs}
\endinsert

If the photon mass $m_{02}\to\infty$ the interaction is switched off, and
so we can use the version (1.3) of the formula for the pressure. With (1.1a) and
(1.5) this reads
$$\eqalign{
P(T)= & P_0^{(1)}(T,m_{01}^2) -\half C
\hbox{ Im }\int _{m_{02}^2}^{\infty}dm_{02}^{\2}
\int {d^{6-2\e}q\over (2\pi)^{6-2\e}}\Big\{
{1+2n(q)\over q^2-m_{02}^{\2}-g_0^2\pi _T (q^0,{\bf q}^2)}\cr
&{}~~~~~~~~~~~~~~~~~~~~~~~~~~~~~~~~~~~~~~~~~~-{1\over
q^2-m_{02}^{\2}-g_0^2\pi (q^2)}
\Big\}}
\eqno(2.1)
$$
where $C=\half\tr\,\btau ^2=3$ is the number of ``photon''
field components in our model.
The integration over the photon mass
$m_{02}'$ is with the coupling $g_0$ and the electron mass
$m_{01}$ fixed. $P_0^{(1)}(T,m_{01}^2)$ is the
contribution to the pressure from the electrons when the interaction is
switched off. The $N$ doublet fields $\psi_r$ correspond to 4$N$
real fields, thus:
$$\eqalign{
P_0^{(1)}(T,m_{01}^2)=&-2N\hbox{ Im }\int _{m_{01}^{2}}^{\infty}dm_{01}^{\2}
\int {d^{6-2\e}q\over (2\pi)^{6-2\e}}\Big\{
{1+2n(q)\over q^2-m_{01}^{\2}}-{1\over q^2-m_{01}^{\2}}\Big\}\cr
=&4N\pi\int {d^{6-2\e}q\over (2\pi)^{6-2\e}}n(q)\theta (q^2-m_{01}^{2})}
\eqno(2.2)
$$
Here, and elsewhere if needed, we assume the usual $i\e$ prescription.

The next step is to express (2.1) and (2.2) in terms of renormalised quantitities.
We introduce a renormalised photon mass $m_2'$ for each value of the bare mass
$m_{02}'$.
The most appropriate scheme is on-shell renormalisation,
in which the renormalised photon mass is given by
$$
m_{2}^{\2}=m_{02}^{\2}+g_0^{2}\Rpi(m_{2}^{\2})
\eqno(2.3)
$$
The electron mass $m_{01}$ is renormalised similarly, though its renormalisation
vanishes as $N^{-1}$ when $N\to\infty$.
%We have chosen to define the renormalised mass to be real, even though
%$\pi(m_{02}^{\2})$ becomes complex for $m_{02}^{\2}>4m_{01}^{\2}$.

We are going to calculate the pressure up to terms of order $N^0$,
so that in the calculation of $\pi (q^2) $  and $\pi _T(q^0,{\bf q}^2)$
we need make no self-energy insertions
in the internal electron lines; these functions  depend just on $q$ and on
the unrenormalised
electron mass $m_{01}$, which we may equate to its
renormalised value $m_1$. However, the renormalisation of $m_{01}$ will
be important below.
To leading order in $N$, the renormalised coupling is
$$
\bar g^2(m_2^{\2})=Z_2(m_2^{\2})g_0^2$$$$
Z_2(m_2^{\2})={1\over 1-g_0^2\Rpi '(m^{\2}_{2})}
={\partial m_2^{\2}\over\partial m^{\2}_{02}}
\eqno(2.4a)
$$
where $\pi '(m^2)$ denotes $\pd\pi (q^2)/\pd q^2$ evaluated at $q^2=m^2$.
Thus, for fixed $g_0^2$,
$$
\bar g^2(m_2^{\2})={g^2\over 1+g^2(\Rpi '(m_2^{2})-\Rpi '(m_2^{\2}))}$$$$
g^2=\bar g^2(m_2^{2})
\eqno(2.4b)
$$
The equations (2.4) give pathologies
reminiscent\ref{\second} of those of
large-$N$ $\phi ^4$ theory. Similar pathologies were discussed
for an exactly solvable model a long time ago\defref\kallen{
G K\"all\'en and W Pauli, Kgl Danske Videnskab Selskab, Mat-Fys Medd
30 (1955) 3 
}.
If we insist that $g_0^2>0$, then we find from
(2.4a) that $g^2\to 0$ as $\e\to 0$. However, if we take the view
that the value of $g_0^2$ is irrelevant for physics, and simply choose
$g^2$ to have some positive value,
we find that $\bar g^2(m_2^{\2})$ has a Landau pole. But provided that
$g^2\ll 192\pi ^3$, the pole is at such a large value $M^2$ of $m_2^{\2}$ that 
it is physically irrelevant: $M^2\sim m_2^{2}\exp (192\pi ^3/g^2)$.
Then our thermal field theory is a sensible theory provided we restrict
ourselves to temperatures $T\ll M$.

The curly bracket in (2.1) is
$$
Z_2(m_2^{\2})\Big\{
{1+2n(q)\over q^2-m_2^{\2}-\bar g^2(m_2^{\2})\bar\pi _T (q^0,{\bf q}^2,m_2^{\2})}
-{1\over q^2-m_2^{\2}-\bar g^2(m_2^{\2})\bar\pi (q^2,m_2^{\2})}
\Big\}
\eqno(2.5)
$$
Here, $\bar\pi _T$ and $\bar\pi$ are convergent functions in 6 dimensions,
when $\e\to 0$:
$$
\bar\pi _T (q^0,{\bf q}^2,m_2^{\2})=\pi _T (q^0,{\bf q}^2)
-(q^2-m_2^{\2})\pi '(m_2^{\2})-\pi (m_2^{\2})$$$$
\bar\pi (q^2,m_2^{\2})=\pi (q^2)
-(q^2-m_2^{\2})\pi '(m_2^{\2})-\pi (m_2^{\2})
\eqno(2.6)
$$
Of course, both $\bar\pi _T$ and $\bar\pi$ depend also on $m_1^2$, but
this is kept fixed at its physical value in both these functions.
We may use (2.4) to change the integration variable from $m_{02}^{\2}$ to
$m_{2}^{\2}$. Because we are taking $g_0^2 < 0$ and $\bar g^2>0$, when
$m_{02}^{\2}$ increases from $m_{02}^{2}$ to $\infty$ we find that $m_{2}^{\2}$
decreases from $m_{2}^{2}$ to $-\infty$. These negative squared masses are
just a calculational device and they do not enter in the final result but,
provided that $m_2 < 2 m_1$,
they mean that the squared renormalised mass
and coupling defined in (2.3) and (2.4) are real throughout the integration,
and the integral in (2.1) is
$$\eqalign{
-\half C
\hbox{ Im }\int _{m_2^2}^{-\infty}dm_2^{\2}
\int {d^{6-2\e}q\over (2\pi)^{6-2\e}}&\Big\{
{1+2n(q)\over q^2-m_2^{\2}-\bar g^2(m_2^{\2})\bar\pi _T (q^0,{\bf q}^2,m_2^{\2})} \cr
&{}~~~~~~~~~~~~~~~~~~~
-{1\over q^2-m_2^{\2}-\bar g^2(m_2^{\2})\bar\pi (q^2,m_2^{\2})}
\Big\}}
\eqno(2.7)
$$
We may perform the mass integration, because from (2.4)
$$
{\partial\over\partial m_2^{\2}}\bar g^2(m_2^{\2})
=\bar g^4(m_2^{\2})\,\pi ''(m_2^{\2})
\eqno(2.8a)
$$
and from (2.6)
$$
{\partial\over\partial m_2^{\2}}\bar\pi _T (q^0,{\bf q}^2,m_2^{\2})=
{\partial\over\partial m_2^{\2}}\bar\pi (q^2,m_2^{\2})=
-(q^2-m_2^{\2})\pi ''(m_2^{\2})
\eqno(2.8b)
$$
From (2.8a), (2.8b)  we find that (2.7) is
$$
C\pi \int {d^{6-2\e}q\over (2\pi)^{6-2\e}} n(q) \theta(q^2-m_2^2)
~~~~~~~~~~~~~~~~~~~~~~~~~~~~~~~~~~~~~~~~~~~~~~~~~~~~~~~~~~~~~~~~~~~~~~~~~~~~$$$$
-\half C \hbox{ Im } \int {d^{6-2\e}q\over (2\pi)^{6-2\e}}
\Bigl[ 2n(q) \log {g^2\bar\pi _T (q^0,{\bf q}^2,m_2^{2})+m_2^2-q^2
\over m_2^{2}-q^2}
+\log {g^2\bar\pi _T (q^0,{\bf q}^2,m_2^{2})+m_2^2-q^2
\over g^2\bar\pi (q^2,m_2^{2})+m_2^2-q^2} \Bigr]
\eqno(2.9)
$$
This integral diverges when $\e\to 0$. The divergence must be cancelled
by a similar one in $P_0^{(1)}(T,m_{01}^2)$, given in (2.2), which we rewrite as
$$
P_0^{(1)}(T,m_{01}^2)=P_0^{(1)}(T,m_{1}^2)+B(T)$$$$
B(T)=-4N\pi\int {d^{6-2\e}k\over (2\pi)^{6-2\e}}n(k)\big [
\theta (k^2-m_1^2)-\theta (k^2-m_{01}^{2})\big ]
\eqno(2.10)
$$
Evidently $P_0^{(1)}(T,m_{1}^2)$ is convergent, but $B(T)$ is not.
Because $m_1^2-m_{01}^2$ goes to zero as $N^{-1}$ as $N$ becomes large,
$$
B(T)\sim
2N(m_1^2-m_{01}^2)\int {d^{6-2\e}k\over (2\pi)^{6-2\e}}
n(k)2\pi \delta (k^2-m_1^2)
\eqno(2.11a)
$$
Figure 1b gives
$$
(m_1^2-m_{01}^2)N=Cig^2\int {d^{6-2\e}q\over (2\pi)^{6-2\e}}
{1\over q^2-m_2^2-g^2\bar\pi(q^2,m_2^2)}
{1\over (k-q)^2-m_1^2}\Big\arrowvert _{k^2=m_1^2}
\eqno(2.11b)
$$
So
$$
B(T)=-\half Cg^2\hbox{ Im }\int {d^{6-2\e}q\over (2\pi)^{6-2\e}}
{1\over q^2-m_2^2-g^2\bar\pi(q^2,m_2^2)}\hat\pi _T(q^0,{\bf q}^2)\eqno(2.11c)
$$$$
\hat\pi _T(q^0,{\bf q}^2)=
4\int {d^{6-2\e}k\over (2\pi)^{6-2\e}}
n(k)2\pi \delta (k^2-m_1^2){1\over (k-q)^2-m_1^2}
\eqno(2.11d)
$$
(We have used the fact that $B(T)$ is real).

We must show that the divergent parts of (2.9) and (2.11c) cancel when
$\e\to 0$.
From the graph of figure 1a,
$$
\pi(q^2)=2i\int {d^{6-2\e}k\over (2\pi)^{6-2\e}}
{1\over k^2-m_1^2}{1\over (q-k)^2-m_1^2}
\eqno(2.12a)
$$
Also
$$
\bar\pi _T ^{11}(q^0,{\bf q}^2,m_2^2)-\bar\pi(q^2,m_2^2)=
{2\over (2\pi)^{6-2\epsilon}}\int d^{6-2\e}k_1\,
d^{6-2\e}k_2\,\delta(k_1+k_2-q)
~~~~~~~~~~~~~~~~~~~~~~~~~~~~~~~~~~~~~~~~~~~~~~~~~~~~~~~~~~~~~~~~~~~~~~~~~~~~~~~~
$$$$~~~~~~
\Big\{
{n(k_1)2\pi\delta(k_1^2-m_1^2) \over k_2^2-m_1^2}+
{n(k_2)2\pi\delta( k_2^2-m_1^2) \over k_1^2-m_1^2}
-in(k_1)n(k_2)2\pi\delta ( k_1^2-m_1^2)2\pi\delta( k_2^2-m_1^2)\Big\}
\eqno(2.12b)
$$
and from this we may calculate $\bar\pi _T(q^0,{\bf q}^2,m_2^2)$
because (1.5c) tells us that
$$
\hbox{Re }\Pi _{rT} = \hbox{Re }\Pi _{rT} ^{11}$$$$
\hbox{Im }\Pi _{rT} = {\hbox{Im }\Pi _{rT} ^{11}\over 1+2n(q)}
\eqno(2.13)
$$
From (2.11d), (2.12b) and (2.13) it is immediate that
$$
\hbox{Re }\hat\pi _T(q^0,{\bf q}^2)=\hbox{Re }
\{\bar\pi _T(q^0,{\bf q}^2,m_2^2)-\bar\pi(q^2,m_2^2)\}
\eqno(2.14a)
$$
It is familiar\defref\elop{
R J Eden, P V Landshoff, D I Olive and J C Polkinghorne, {\it The analytic
$S$-matrix}, Cambridge University Press (1966)
}
that Im $\pi (q^2)$ is even in $q^0$ and
$$
\hbox{Im }\pi(q^2)=-\theta(q^2-4m_1^2)
{1\over (2\pi)^{6-2\e}}\int d^{6-2\e}k_1d^{6-2\e}k_2\delta(k_1+k_2-q)
2\pi\delta ( k_1^2-m_1^2)2\pi\delta( k_2^2-m_1^2)
\eqno(2.15a)
$$
From (2.12b) we see that\goodbreak
$$
\hbox{Im }\{\bar\pi _T ^{11}(q^0,{\bf q}^2,m_2^2)-\bar\pi(q^2,m_2^2)\}=
~~~~~~~~~~~~~~~~~~~~~~~~~~~~~~~~~~~~~~~~~~~~~~~~~~~~~~~~~~~~~~~~~~~~~~~~~~~~~~~~
$$$$
~~~-{1\over (2\pi)^{6-2\e}}\int d^{6-2\e}k_1d^{6-2\e}k_2\delta(k_1+k_2-q)
2\pi\delta ( k_1^2-m_1^2)2\pi\delta( k_2^2-m_1^2)
\{n(k_1)+n(k_2)+2n(k_1)n(k_2)\}
\eqno(2.15b)
$$
This is nonzero for both $q^2>4m_1^2$ and $q^2<0$.
For $q^2>4m_1^2$ the $\delta$-functions demand that both $k_1^0$
and $k_2^0$ are positive or both negative, so that
$$
n(k_1)+n(k_2)+2n(k_1)n(k_2)=\{1+2n(q)\}\{1+n(k_1)+n(k_2)\}-1
\eqno(2.16a)
$$
and we find that the relation (2.14a) is true also for the imaginary parts
of the functions involved. But for $q^2<0$ the $\delta$-functions require
$k_1^0$ and $k_2^0$ to have opposite signs, and instead
$$
n(k_1)+n(k_2)+2n(k_1)n(k_2)=\e (q^0)\e(k_1^0)\{1+2n(q)\}\{n(k_2)-n(k_1)\}
\eqno(2.16b)
$$
In consequence,
$$
\hbox{Im }\hat\pi _T(q^0,{\bf q}^2)=\hbox{Im }
\{\bar\pi _T(q^0,{\bf q}^2,m_2^2)-\bar\pi(q^2,m_2^2)\}
- \theta(-q^2)\phi _T(q^0,{\bf q}^2)$$$$
\phi _T(q^0,{\bf q}^2)=\theta(q^0)\hat\phi _T(q^0,{\bf q}^2)
+\theta(-q^0)\hat\phi _T(-q^0,{\bf q}^2)$$$$
\theta(q^0)\hat\phi _T(q^0,{\bf q}^2)=4\theta(q^0)
{1\over (2\pi)^{6-2\e}}\int d^{6-2\e}k_1d^{6-2\e}k_2\delta(k_1+k_2-q)
2\pi\delta ( k_1^2-m_1^2)2\pi\delta( k_2^2-m_1^2)
\theta (k_1^0)n(k_1)
\eqno(2.14b)
$$

So finally, when the free-field pressure is subtracted off,
the interaction pressure is
$$\eqalign{
P(T)^{{\rm int}}&=
-\half C \hbox{ Im } \int {d^{6-2\epsilon}q\over (2\pi)^{6-2\epsilon}}
\Bigl[ 2n(q) \log {g^2\bar\pi _T (q^0,{\bf q}^2,m_2^{2})+m_2^2-q^2
\over m_2^{2}-q^2} \cr
&\qquad +\log {g^2\bar\pi _T (q^0,{\bf q}^2,m_2^{2})+m_2^2-q^2
\over g^2\bar\pi (q^2,m_2^{2})+m_2^2-q^2}
+g^2 {\bar\pi _T (q^0,{\bf q}^2,m_2^{2})-
\bar\pi (q^2,m_2^{2})-i\theta(-q^2)\phi _T(q^0,{\bf q}^2)\}\over
q^2-m_2^{2}-g^2\bar\pi (q^2,m_2^{2})} \Bigr] }
\eqno(2.17)
$$
For $\epsilon\to0$, that is in 6 dimensions,
the integrand in (2.17) is finite, so it remains to check that its high-$q$
behaviour is such that the integral is finite.
From (2.12a)
$$
\pi(q^2)=-{2\over(4\pi)^{3-\e}}\Gamma(-1+\e )\int _0^1dx\, [m_1^2-q^2x(1-x)]^{1-\e}
\eqno(2.12c)
$$
so that
$$
\bar\pi (q^2,m_2^2)=-{2\over(4\pi)^{3-\e}}\Gamma(-1+\e )\int _0^1dx\Big\{
[m_1^2-q^2x(1-x)]^{1-\e}-[m_1^2-m^2_2x(1-x)]^{1-\e}\,+
~~~~~~~~~~~~~~~~~~~~~~~~~~~~~~~$$$$
~~~~~~~~~~~~~~~~~~~~~~~~~~~~~~~~~~~~~~~
~(q^2-m^2_2)x(1-x)(1-\e )[m_1^2-m_2^2x(1-x)]^{-\e}\Big\}
\eqno(2.12d)
$$
When $q^2$ becomes large, $\bar\pi (q^2,m_2^2)$ contains terms of order
$q^2$ and $q^{2-2\e}$, while the integral shown in (2.12b) that gives the
difference between $\bar\pi _T ^{11}$ and $\bar\pi$ is only of order
$1/q^2$, so that $\bar\pi _T(q^0,{\bf q}^2,m_2^2)\sim\bar\pi (q^2,m_2^2)$
and
$$
\log {g^2\bar\pi _T (q^0,{\bf q}^2,m_2^{2})+m_2^2-q^2
\over g^2\bar\pi (q^2,m_2^{2})+m_2^2-q^2}
\sim -g^2{\bar\pi _T (q^0,{\bf q}^2,m_2^{2})-\bar\pi (q^2,m_2^{2})
\over q^2-m_2^{2}-g^2\bar\pi (q^2,m_2^{2})}
+O(1/q^8)
\eqno(2.18)
$$
On the other hand one can see from (2.14b) that when $|{\bf q}|$ is large,
whether or not $q^0$ also is large, $\phi _T(q^0,{\bf q}^2)$
is exponentially small. Hence when $q^2$ is large the last two terms in the
integrand of (2.17) combine to make the integral over $q$ convergent.

The central result of this section is our formula (2.17) for the
interaction pressure to order $N^0$. The reader may worry about some of the
steps in its derivation, for instance
an integration over negative renormalised squared masses for the
photon in (2.7). Thus in intermediate steps we considered tachyonic
photons! In the next section we will discuss the general
renormalisation problem for the pressure. This will lead us in
the $1/N$ expansion to another derivation of (2.17) which avoids
these problems.
\bigskip\goodbreak
{\bf 3 Renormalisability of the pressure}

We now return to the interaction (1.6) and show how it leads in 6 dimensions
to a finite expression for the pressure even when we do not take the large-$N$ 
limit. The form (1.6) is designed to simulate QED without introducing the
complications that arise from spin, and so our analysis follows closely
the familiar renormalisation of zero-temperature QED,
such as is described in the book of Bjorken and
Drell\defref\bd{
J D Bjorken and S D Drell, {\sl  Relativistic quantum fields}, McGraw-Hill
(1965)
}.
As we shall see, the task of expressing the derivatives of the pressure
with respect to the masses in terms of renormalised propagator and vertex
functions leads us to a problem of overlapping divergences. This turns out to
be similar to the overlapping-divergence problem for the vacuum polarisation
in QED, where, following Dyson, one introduces a certain electron-positron
scattering kernel (see chapter 19 of [\bd]). We shall follow a similar road
here.

Our notation will be as follows. We use the labels $\alpha,\beta,\dots$
and $a,b,\dots$ to denote components of isodoublets and isotriplets,
respectively, and $r,s,\dots$ for flavour labels. We also write the
unrenormalised fields of the ``electrons'' and ``photons'' together as
$$
\phi _{1r}=\psi _{r}$$$$
\phi _{2}^a=A^a
\eqno(3.1)
$$
These are the unrenormalised fields. As in the last section, their
unrenormalised masses are $m_{0i}$, \hbox{$\,i=1,2$}. Where it does not cause
confusion, we will not explicitly write the isospin and flavour labels.
The renormalised fields will be labelled with an
additional suffix \R; their masses are $m_{i}$, which are related to the
unrenormalised masses by
$$
m_i^2-m_{0i}^2-\Pi_i(q^2)\big\arrowvert _{q^2=m_i^2}=0~~~~~~~~~~~~~~~~~~~~~i=1,2
\eqno(3.2a)
$$
Here, $\Pi_1(q^2)$ and $\Pi_2(q^2)$ are the self energies with the various
Kronecker deltas factored off.  From them, we also construct the two
wave-function renormalisation constants
$$
Z_i=\left [1-{\pd\Pi_i(q^2)\over\pd q^2}\right ]^{-1}_{q^2=m_i^2}
\eqno(3.2b)
$$
and hence the renormalised fields
$$
\phi_{i\R}=Z_i^{-1/2}\phi_i
\eqno(3.2c)
$$
We are assuming 
that $0\leq m_2^2 <4m_1^2$, so that the renormalisations
are real.
The propagators before and after renormalisation  are
$$
D _{i}(q^2)=i[q^2-m^2_i-\Pi_{i}(q^2)]^{-1}\cdot\eins$$$$
D _{i\R}(q^2)=Z_i^{-1}D _{i}(q^2)=i[q^2-m^2_i-\bar{\Pi}_{i}(q^2)]^{-1}\cdot\eins
\eqno(3.2d)
$$
where the two functions
$$
\bar{\Pi}_{i}(q^2)=Z_i[\Pi_i(q^2)-\Pi_i(m_i^2)]-(Z_i-1)(q^2-m_i^2)
~~~~~~~~~~~~~~~~~~~~~i=1,2
\eqno(3.2e)
$$
are finite and have value zero and zero derivative at $q^2=m_i^2$.
In (3.2d) $\eins$ is to be read as $\delta_{\alpha\beta}
\delta_{rs}$ for $i=1$ and as $\delta_{ab}$ for $i=2$.

We introduce the one-particle-irreducible vertex function $\gamma (p',p)$
which couples a pair of electrons of momenta $p,p'$ to a photon. Renormalise
it and the coupling, so that
$$
\lambda \gamma_\R(p',p)=\lambda_0Z_1Z_2^{1/2}\gamma (p',p).
\eqno(3.3a)
$$
We choose to fix $\gamma_\R(p',p)$, and so define the renormalised coupling
$\lambda$, by imposing the condition
$$
\gamma_\R\Big\arrowvert_{\rm SP}=1,\eqno(3.3b)$$
where
$$
\gamma_\R\Big\arrowvert_{\rm SP}= \gamma_\R(p, p')\qquad
{\rm evaluated\ for\ }\quad p^2=p^{'2} = (p-p')^2=-M^2
\eqno(3.3c)
$$
for some fixed mass $M$.
This renormalisation absorbs, for example,
 a divergence from the one-loop triangle graph
(which we could neglect in the large-$N$ limit).

\def\Z{{\bf Z}}
\def\T{{\bf T}}
\def\P{{\bf P}}
\def\K{{\bf K}}
Note that the coupling (1.6) is invariant under the following
$C$-transformation
$$\eqalign{
C:\quad &\phi_{1r}\longrightarrow \epsilon(\phi^\dagger_{1r})^T\quad(r=1,...,N),\cr
&\phi_2\longrightarrow-\phi_2\cr}\eqno(3.3d)$$
where
$$\epsilon=\pmatrix{0&1\cr
-1&0\cr}.\eqno(3.3e)$$
This forbids a nonzero vertex function for three
``photons'' in our model, similarly to Furry's theorem in QED.
By the usual power counting arguments (compare [6]), we see then
that the mass, wave function and coupling renormalisations (3.2), (3.3a)--(3.3c)
make the (perturbative) theory finite.

We need to introduce a $2\to 2$-body connected scattering amplitude \T.
This is a $2\times 2$ matrix connecting the channels
\smallskip
\settabs 4\columns
\+&1:$~~~~~~~\phi_1 +\bar\phi_1$&\cr
\+&2:$~~~~~~~\phi_2+\phi_2$&\cr
\vskip -10pt\rightline{(3.4)}\vskip -5pt
Each element of \T\ is also a matrix in isospin and flavour space.
We define \T\ to be amputated --  there are no single-particle poles in its
external legs -- and we exclude from it all terms which are one-particle
reducible in the $s$-channel,
though $T_{11}$ does have a $t$-channel photon pole and $T_{12}$ and $T_{21}$
have $t$ and $u$ channel electron poles.
We also introduce the two-particle-irreducible kernel \K\
associated with \T. It has no $s$-channel two-particle intermediate states
and is related to \T\ by
$$
\T=\K+\K\P\T=\K+\T\P\K
\eqno(3.5a)
$$
Here the $2\times 2$ matrix \P\ is diagonal; one diagonal
element $P_{11}$ is
the tensor product of two electron propagators, and the
other $P_{22}$ is half the tensor product of two photon propagators. The
{\eightrm 12} element of the first matrix equation in (3.5a) is drawn in
figure 2, together with the definition of the matrix \P.
The factor $\half$
in the definition of \P\ takes account of the symmetry of channel 2 under
interchange of the two photons. Because of the relation (3.2d) between
the renormalised and unrenormalised propagators,
$$
\P _{\R}=\Z ^{-1}\P\Z ^{-1}
$$$$
\Z =\left [\matrix{Z_1&0\cr
               0&Z_2\cr}\right ]
\eqno(3.5b)
$$
On the other hand, an amputated scattering amplitude on renormalisation
acquires a factor
$Z^{1/2}$ for each external leg, so the renormalised versions of \T\ and \K\
are
$$
\T _{\R}=\Z\T\Z ~~~~~~~~~~~~~~~~~~~~~~~\K _{\R}=\Z\K\Z
\eqno(3.5c)
$$
Hence after renormalisation
$$
\T_{\R}=\K_{\R}+\K_{\R}\P_{\R}\T_{\R}=\K_{\R}+\T_{\R}\P_{\R}\K_{\R}
\eqno(3.5d)
$$

Note that $\T$ and $\K$ have skeleton expansions that express them in a unique
way in terms of integrals over the unrenormalised propagators and vertex
functions in (3.2) and (3.3), while $\T_{\R}$ and $\K_{\R}$ have similar
expansions in terms of the renormalised functions.
\topinsert
\centerline{{\epsfxsize=100mm\epsfbox{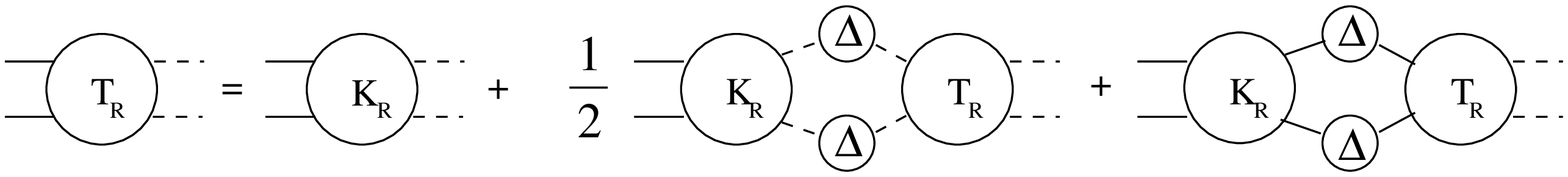}}}
\vskip 5mm
\centerline{{\epsfxsize=50mm\epsfbox{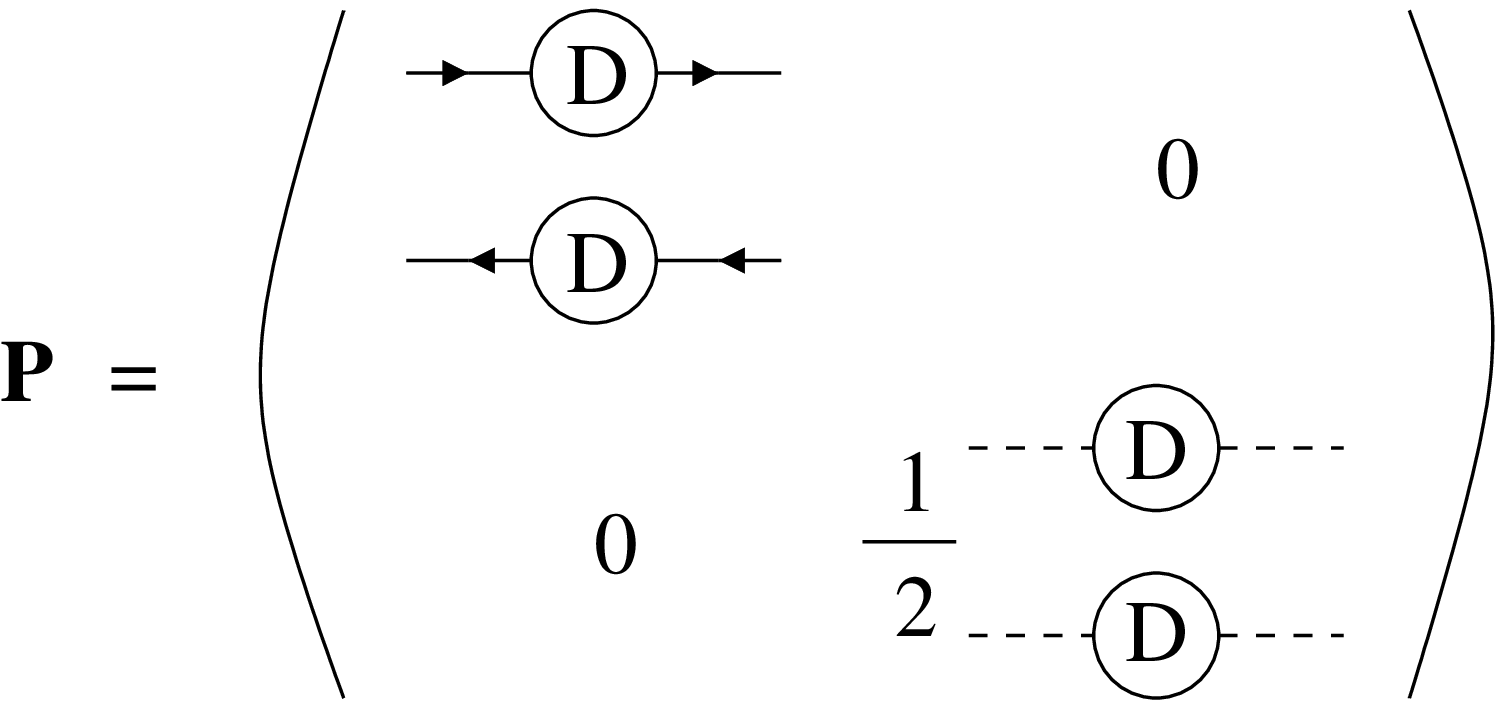}}}
\vskip 4pt
\centerline{Figure 2: one of the equations (3.5a) and the definition of the
propagator matrix \P}
\vskip 7mm
\leftskip4mm\rightskip4mm
\centerline{{\epsfxsize=0.8\hsize\epsfbox{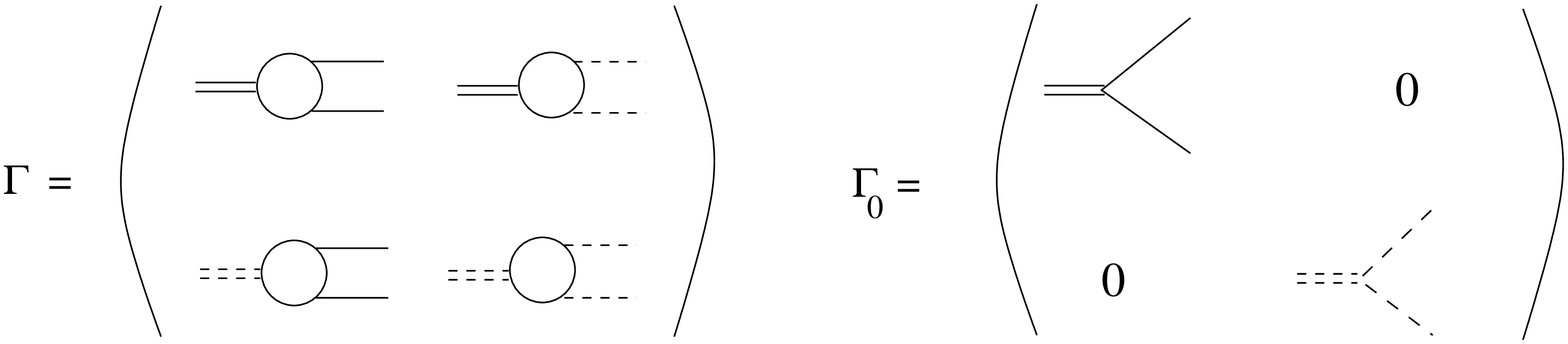}}}
\vskip 1pt
Figure 3: Definition of the matrix of Green's functions that
couple the composite fields to the electron-positron and photon-photon
channels (3.4)
\endinsert

We now define the composite operators constructed from the
electron and photon fields
$$\eqalignno{
\chi _1(x)&=\phi_1^{\dag}(x)\phi_1(x)-\langle
0|\phi_1^{\dag}(x)\phi_1(x)|0\rangle\cr
\chi _2(x)&=\half [\phi^2_2(x)-\langle 0|\phi^2_2(x)|0\rangle ]
&(3.6)\cr}
$$
Here, a sum over flavour and isospin indices is implied.
Define also the one-particle
irreducible Green's functions that couple these to the channels 1 and 2
defined in (3.4);
they form a $2\times 2$ matrix ${\bf \Gamma}$, shown graphically in figure 3.
Each element of the matrix is a function of the momenta $p$ and $p'$ of the
final-state particles. In fact for our purposes we need only consider the
case $p=p'$, so that ${\bf \Gamma}$ is a function just of the single variable
$p^2$.
To zeroth order in perturbation theory, ${\bf \Gamma}={\bf\Gamma}_0$ which,
apart from Kronecker deltas for flavour and isospin indices is just the unit
matrix:
$$
{\bf\Gamma}_0=\left (\matrix{\delta_{rs}\delta_{\alpha\beta}&0\cr
                             0&\delta_{ab}\cr}\right )
\eqno(3.7a)
$$
The complete ${\bf\Gamma}$  has flavour and isospin structure
\def\G{{\hat\Gamma}}
$$
{\bf\Gamma}=\left (\matrix{\G _{11}(p^2)\delta_{rs}\delta_{\alpha\beta}&
        \G _{12}(p^2)\delta _{ab}\cr
\G _{21}(p^2)\delta_{rs}\delta_{\alpha\beta}&\G _{22}(p^2)\delta_{ab}\cr}
\right )
\eqno(3.7b)
$$
and it satisfies a Dyson equation
$$\eqalignno{
{\bf \Gamma}&={\bf \Gamma }_0+{\bf \Gamma }_0\P\T \cr
            &={\bf \Gamma }_0+{\bf \Gamma }\P\K &
(3.7c)\cr}
$$
where we have used (3.5a). We define renormalised composite fields by
$$
\bchi _{\R}=\Z _{\chi} ^{-1}\bchi
\eqno(3.8a)
$$
where the matrix $\Z _{\chi}$ is to be chosen; then
$$
{\bf \Gamma}_{\R}=\Z _{\chi}^{-1}{\bf \Gamma}\Z
=\Z _{\chi}^{-1}{\bf \Gamma }_0\Z+{\bf \Gamma} _{\rm R}\P _{\rm R}\K _{\rm R}
\eqno(3.8b)
$$
The isospin and flavour structure of  ${\bf \Gamma}_{\R}$ is the same as that
of ${\bf \Gamma}$, that is (3.7b) but with scalar functions $\G _{ij{\rm R}}(p^2)$.
Using the usual power-counting argument
and the skeleton expansion
for $\K _{\rm R}$ we see that indeed ${\bf \Gamma}_{\R}$ can be made finite by
suitable choice of $\Z _{\chi}$, in such a way that the divergence
from the last term in (3.8b) is absorbed in each order of
perturbation theory (compare for example chapter 19 of [{\bd}]).
We also see from (3.8b) that any $\Z _{\chi}$ that makes
the elements of ${\bf \Gamma}_{\R}$
finite at one (but not necessarily the same)
point in momentum space is an acceptable choice for this
renormalisation matrix.

In order to choose $\Z _{\chi}$ appropriately,
we return to the relation (3.2a) between the
renormalised and unrenormalised masses. Remembering that $\Pi _i(q^2)$
depend also on $m_{01}^2$ and $m_{02}^2$, we have
$$
{\pd m_j^2\over\pd m_{0i}^2}-\delta _{ij}-
{\pd\over\pd m_{0i}^2}\Pi_j(q^2)\Big\arrowvert _{q^2=m_j^2}
-{\pd\over\pd q^2}\Pi_j(q^2)\Big\arrowvert _{q^2=m_j^2}{\pd m_j^2\over\pd
m_{0i}^2}=0
\eqno(3.9a)
$$
This gives, with the definition (3.2b) of $Z_i$,
$$
{\pd m_j^2\over\pd m_{0i}^2}Z_j^{-1}=\delta _{ij}+
{\pd\over\pd m_{0i}^2}\Pi_j(q^2)\Big\arrowvert _{q^2=m_j^2}
\eqno(3.9b)
$$
The differentiation with respect to $m_{0i}^2$
is with fixed bare coupling $\lambda _0$.
When applied to any Feynman graph for $\Pi _j$, it gives a sum of terms
in which each internal line of type $i$ in turn is doubled. We recognise
therefore that the quantity on the right-hand side of (3.9b) is just
$$
\delta _{ij}+{\pd\over\pd m_{0i}^2}\Pi_j(q^2)=\G _{ij}(q^2)
\eqno(3.9c)
$$
evaluated at $q^2=m_j^2$. Hence, if we define
$$
[\Z _{\chi}]_{ ij}={\pd m_j^2\over\pd m_{0i}^2}
\eqno(3.9d)
$$
we see from (3.5b) and (3.8b) that
$$
\G _{ij\R}(q^2)\Big\arrowvert _{q^2=m_j^2}= \delta_{ij}
\eqno(3.10a)
$$
Therefore the elements of ${\bf \Gamma}_{\R}(q^2)$ are finite when the
final-state particles $q$ are on shell,
that is at one point in momentum space. From the arguments
given above it follows that they are also
finite when the particles are off shell.

We are going to calculate the derivative of the pressure with
respect to the renormalised masses at fixed bare coupling. It is
convenient to introduce a renormalisation prescription in
which, unlike in (3.2) and (3.3), fixed bare coupling corresponds
to fixed renormalised coupling while the renormalised masses vary.
In this modified prescription we replace the bare masses $m_{0i}$
in (3.2a) with fixed bare masses $\mu_{0i}$, though keep the same bare
coupling $\lambda_0$. Then (3.2b), (3.2c) and (3.3) lead to
a modified renormalised coupling $\tilde\lambda$. We claim that
$\tilde\lambda$ is a finite function of $\lambda$. To show this,
differentiate the definition (3.3) of $\lambda$ with respect
to the bare masses at fixed $\lambda_0$:
$$
{\partial\lambda\over\partial m^2_{0j}}=\lambda_0
Z_1Z_2^{1\over2}{\partial\gamma(p',p)|_{\rm SP}\over \partial m^2_{0j}}
+\lambda{\partial\over \partial m^2_{0j}}\log
Z_1+{1\over2}\lambda{\partial\over\partial m^2_{0j}}\log Z_2
\eqno(3.11a)$$
Now
$$\lambda_0{\partial\gamma(p',p)\over \partial m^2_{0j}}
=\gamma_j(p',p)\eqno(3.11b)$$
is the unrenormalised amputated Green's function for the fields
$\chi_j,\phi^\dagger_1,\phi_1, \phi_2$ at zero four momentum
for the $\chi_j$-field. Its superficial degree of divergence
is $-2$, so it is rendered finite by the renormalisation
$$\gamma_j(p',p)Z_1Z_2^{1\over2}=\sum_{k=1,2}[\Z_\chi]_{jk}\,
\gamma_{k{\rm R}}(p',p)\eqno(3.11c)$$
Differentiating the definition (3.2b) of the $Z_i$ gives
$$
{\partial\over \partial m^2_{0j}}\log
Z_i=\sum_{k=1,2}\left.\left\{\delta_{ki}{\partial^2\bar
\Pi_i(q^2)\over(\partial q^2)^2}+{\partial\over\partial q^2}\hat\Gamma_
{ki{\rm R}}(q^2)\right\}\right|_{q^2=m^2_i}{\partial m^2_k\over
\partial m^2_{0j}}\eqno(3.11d)$$
Then (3.11a), together with (3.9d), gives
$$\eqalign{
{\partial\lambda\over\partial m^2_1}&=\gamma_{1{\rm R}}(p',p)|_{{\rm SP}}
+\lambda\left.
\left\{{\partial^2\bar\Pi_1(q^2)\over(\partial q^2)^2}
+{\partial\over \partial q^2}\hat\Gamma_{11{\rm R}}(q^2)\right\}\right|
_{q^2=m^2_1}+\half\lambda{\partial\over\partial q^2}\hat\Gamma
_{12{\rm R}}(q^2)|_{q^2=m^2_2}\cr
{\partial\lambda\over\partial m^2_2}&=\gamma_{2{\rm R}}(p',p)|_{{\rm SP}}
+\lambda{\partial\over \partial q^2}\hat
\Gamma_{21{\rm R}}(q^2)|_{q^2=m^2_1}+\half\lambda\left.
\left\{{\partial^2\bar\Pi_2(q^2)\over(\partial q^2)^2}
+{\partial\over \partial q^2}\hat\Gamma_{22{\rm R}}(q^2)\right\}\right|
_{q^2=m^2_2}\cr}\eqno(3.11e)$$
This expresses, in terms of renormalised quantities only,
how $\lambda$ changes when the renormalised masses are changed,
keeping $\lambda_0$ fixed. Thus, as we have claimed above, two
renormalised couplings for different renormalised masses but the
same $\lambda_0$ differ by a finite amount. The definition
(3.3a,b) of $\lambda$ expresses it as a function of
$\lambda_0$ and the bare masses $m_{0i}$, which in turn
can be considered from (3.2a) as functions of $\lambda_0$
and the renormalised masses $m_i$. Then from $\lambda$ considered
as function of $\lambda_0$ and the masses $m_i$ and the corresponding
definition of $\tilde\lambda$ as function of $\lambda_0$
and the masses $\mu_i$ we get two equations from which, in
principle, we may eliminate $\lambda_0$ and so express $\tilde
\lambda$ as a function of $\lambda$ (though in practice this will
be a nontrivial task). However, from its definition it is clear
that $\tilde\lambda$ remains fixed when the masses $m_i$ vary
in differentiations with fixed $\lambda_0$.

We now return to formula (1.1). We choose the version (1.1b) and
write it in the form
$$
{\partial\over\partial m^2_{0i}}
P(T)=-\sum_j\int {d^nq\over (2\pi )^n}
{\tr}\left[\Gamma_{0ij}
2^{1-j}\Delta^{11}_{j T}(q)\right]
%\Delta _{iT}(q)
~~~~~~~~~~~~~~~~i=1,2
\eqno(3.12a)
$$
Here ${\bf \Gamma}_0$ is defined in (3.7a) and
$$\Delta^{11}_{jT}(q)=D^{11}_{jT}(q)-D_j(q^2)
= \hat{\Delta}^{11}_{jT}(q)\cdot\eins \eqno(3.12b) $$
is the difference of the unrenormalised thermal 11-propagator and
the zero temperature propagator for the field $\phi_j
\ (j=1,2)$. The trace is with respect to the isospin
and flavour indices.

\topinsert
\centerline{{\epsfxsize=100mm\epsfbox{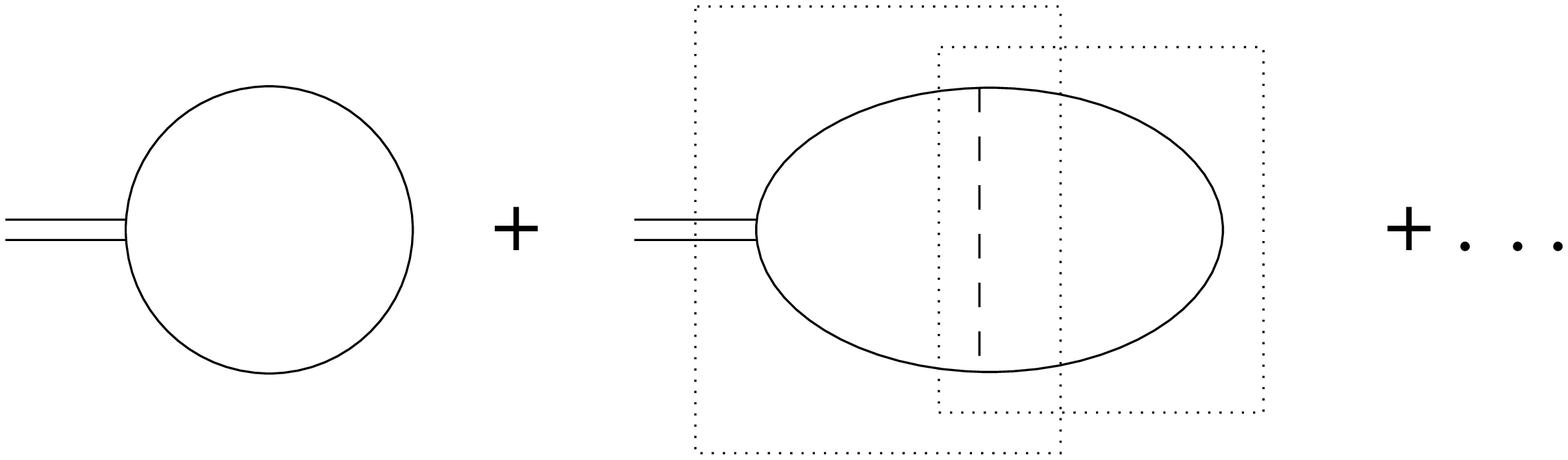}}}
\vskip 1pt
\centerline{Figure 4: The diagrammatic expansion (3.12a)
for $\partial P(T)/\partial m^2_{01}$ }
\endinsert
The renormalisation of (3.12a) leads to a problem of
overlapping divergences. To see this we consider 
the diagrammatic expansion of $\partial P(T)/\partial
m_{01}^2$, for example, shown  in figure 4.
The second term of figure 4  can be interpreted as a correction
either to the propagator on the right or to the vertex
function of the composite operator $\chi_1$ on the left.
This is analogous to the case of the vacuum polarisation
function in QED (compare chapter 19 of [{\bd}]),  as will be our methods
for a proper renormalisation of (3.12a).

We express $\Delta _{iT}^{11}$ in terms of $\Delta _{iT\R}^{11}
$, which is defined in a
similar way, but with the renormalised fields replacing the bare fields
so that
$$
\Delta^{11}_{iT{\rm R}}(q)=Z_i^{-1}\Delta^{11}_{iT}.\eqno(3.12b)$$
%Just as in (25),
%$$\eqalignno{
%\Delta _{iT\R}^{11}(q)&=Z_i^{-1}\Delta^{11} _{iT}\cr
% &=\hbox{ Im }
%\Big\{{1+2n(q)\over q^2-m_{i}^{2}-\Pi _{iT\C} (q^0,{\bf q}^2)}-{1\over
%q^2-m_{i}^{2}-\Pi _{i\C} (q^2)}
%\Big\}&(3.11b)\cr}$$
%where
%$$
%\Pi_{iT\C}(q^2)=Z_i[\Pi_{iT}(q^2)-\Pi_i(m_i^2)]-(Z_i-1)(q^2-m_i^2)
%\eqno(3.11c)$$
%is finite.
We use the fact that ${\bf\Gamma}_0$ is a constant matrix,
but can be expressed from (3.7c) as
$$
{\bf \Gamma}_0={\bf \Gamma}- {\bf \Gamma P K}\eqno(3.12ba)$$
where the individual terms on the right-hand side are momentum-dependent.
Now we choose as momentum argument on the right-hand side just the
integration variable $q$ of (3.12a). With this
and the definition (3.9d) of the matrix $[\Z _{\chi}]_{ ij}$, we
have
$$
{\partial\over\partial m^2_{i}} P(T)=-\sum _j
\int {d^nq\over (2\pi )^n}{\tr}\Big\{
\Big [ \Z^{-1}_{\chi}\big ({\bf\Gamma}-{\bf\Gamma}\P\K\big )\Z)\Big ]_{ij}
2^{1-j}\Delta^{11}_{jT\R}\Big\}
\eqno(3.12c)
$$
or
$$
{\partial\over\partial m^2_{i}}P(T)=-\sum _j
\int {d^nq\over (2\pi )^n}{\tr}\Big\{
\Big [\big ({\bf\Gamma_\R}-{\bf\Gamma_\R}\P _\R\K _\R\big )\Big ]_{ij}
2^{1-j}\Delta^{11} _{jT\R}\Big\}
\eqno(3.12d)
$$
In using the formula, we must remember that the partial derivative is
with the bare coupling $\lambda _0$ fixed, so
the appropriate renormalised coupling is $\tilde\lambda$.

\topinsert
\centerline{{\epsfxsize=100mm\epsfbox{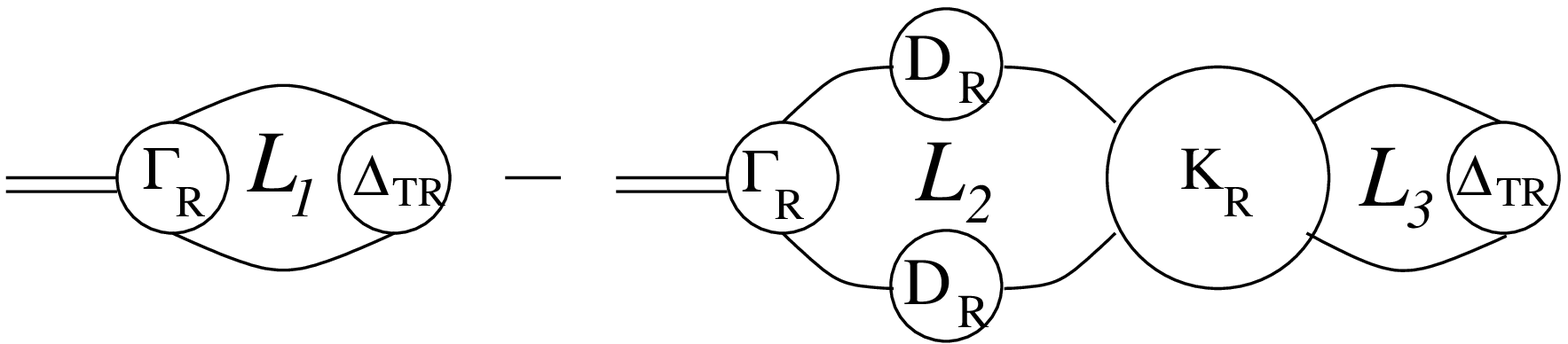}}}
\vskip 1pt
\centerline{Figure 5: The expression on the right-hand side of (3.12d)}
\endinsert

The right-hand side of the expression (3.12d) is shown graphically in
figure 5. The renormalised vertex, kernel and propagator functions
$\Gamma_{\rm R},\, K_{\rm R},\ D_{\rm R},\ \Delta_{T{\rm R}}^{11}$ are, of course,
finite for $n=6$. But we still have three loop integrals to do, which we
call $L_1,L_2$ and $L_3$, as indicated  in figure 5.

Let us first determine the degree of divergence for $n=6$ of
the loop integrations $L_1$ and $L_3$.
To this end, we write an operator-product expansion:
$$
\phi_{i\R}(x)\phi_{i\R}^{\dag}(0)=C_i(x^2)1\!\!1 +\sum_jC_{ij}(x^2)
\,\chi _{j\R}(0)+\dots
\eqno(3.13a)
$$
 As
the $\chi _{j\R}$ have the same dimension as
the left-hand side of (3.13a),
the small-distance behaviour of $C_{ij}(x^2)$ is
$(x^2)^0$, up to possible logarithms.
The thermal propagator $D^{11}_{iT{\rm R}}(q)$ is the Fourier
transform of the vacuum expectation value of (3.13a).
Because $\Delta^{11}_{iT{\rm R}}(q)$ is the difference between the
temperature $T$ and temperature 0 propagators, the first
term on the right-hand side of (3.13a) does not contribute
to it and the leading power behaviour for $q\to\infty$ is
obtained from the second term on the right-hand side of (3.13a).
Thus, at worst, 
$$
\Delta _{iT\R}(q)\sim (q^2)^{-3} \quad {\rm for}\quad q\to\infty.
\eqno(3.13b)
$$
Power counting shows that the $L_1$ integration is logarithmically divergent.
For the second term in figure 5, we choose to do the loop integration 
$L_3$ first. Superficially, it is convergent. It really does
converge if there are no
divergent subintegrations. Of course,
no divergent subintegrations occur in $K_{\rm R}$ and $\Delta_{T{\rm R}}$.
Thus we could get a divergent subintegration only\defref\iz{
C Itzykson and J-B Zuber, {\it Quantum field theory}, McGraw-Hill (1980)
chapter 8\h
S Weinberg, Physical Review 118 (1960) 838
}
if the loop $L_3$
is closed in $K_{\rm R}$ directly on one single skeleton vertex. But this
cannot happen because
$K_{\rm R}$ contains no $s$-channel 1-particle-reducible
diagrams. Thus we conclude that the $L_3$
loop is convergent; it gives a high-$q^2$ behaviour $ 1/q^2$ to that loop.
From this, we see that the $L_2$ integration diverges
logarithmically.

So both terms in figure 5
are logarithmically divergent. It remains to show that these
divergences cancel, leaving a finite result. From the discussion
above it is clear that we will have to consider only the part
of $\Delta_{iT{\rm R}}(q)$ proportional to $(q^2)^{-3}$ for
$q\to\infty$. The higher terms in the operator product
expansion (3.13a) lead to
convergent contributions in all loops $L_1,\ L_2,\ L_3$. Thus,
for the discussion of the convergence we can replace
$\Delta^{11}_{iT{\rm R}}$ in (3.12d) and figure 5 by
any expression having the same $(q^2)^{-3}$ behaviour for $q\to\infty$.
We choose the following 4-point Green's functions:
\def\TT{{\rm T}}
$$\eqalign{
J_{4ik}(q,p)
=\int d^nx\int d^nz_1\int d^nz_2\,&e^{iqx}e^{-ipz_1}
e^{ipz_2}
 \Big\{\langle 0|\TT(\phi_{i{\rm R}}(x)\phi^\dagger_{i{\rm R}}(0)\phi^\dagger_{k{\rm R}}
(z_1)\phi_{k{\rm R}}(z_2))|0\rangle \cr
&-\langle 0|\TT(\phi_{i{\rm R}}(x)\phi^\dagger_{i{\rm R}}(0))|0\rangle 
\langle 0|\TT(\phi^\dagger_{k{\rm R}}(z_1)\phi_{k{\rm R}}(z_2))|0\rangle\Big\}
\cr}\eqno(3.16)$$
where $p$ is an arbitrary fixed momentum. For simplicity, we continue
not to write explicitly the 
isospin and flavour indices
of the fields.  Those attached to $\phi_{i{\rm R}}$ and $\phi^\dagger_{i{\rm R}}$ are
carried by $J_{4ik}$, while those attached to $\phi_{k{\rm R}}$ and
$\phi^\dagger_{k{\rm R}}$ are equal to each other and summed.
This summation excludes $s$-channel one particle reducible
diagrams from $J_{4ik}$.
Inserting here the operator product
expansion (3.13a), we see that the leading term 
for $q\to \infty$ is
again given by the $C_{ij}(x^2)$. In the loop integrals $L_{1,2,3}$ with the
thermal propagators $\Delta_{iT{\rm R}}$ possible divergences
can only be proportional to the two thermal expectation
values of $\chi_{j{\rm R}}(0)$ in the operator product
expansion (3.13a), where $j=1,2$.
Thus, in order to prove convergence, we choose two ``trial''
terms $J_4$ corresponding to the index $k=1,2$ in (3.16) which have
linearly independent contributions from the
$\chi_{j{\rm R}}(0)$ $(j=1,2)$ in the operator product
expansion. The structure of $J_{4ik}$
is as follows (compare (3.5c), (3.5d) and figure 6):
$$\eqalign{
J_{4ik}(q,p)={\tr}_p\Big\{&
2^{i-1}\left[\P_{\rm R}(p)\right]_{ik}
(2\pi)^n[\delta(q-p)+\delta_{k2}\delta(q+p)]+
\cr &
2^{i-1}\left[\P_{\rm R}(q){\bf T}_{\rm R}(q,p)\P_{\rm R}(p)\right]_{ik}2^{k-1}
\Big\}}
\eqno(3.17)$$
%\cr
%&{}
%\eqno(3.17)$$
where ${\tr}_p$ stands for the summation over the isospin and
flavour quantum numbers connected with the $p$-lines in figure 6.
\topinsert
\centerline{{\epsfxsize=130mm\epsfbox{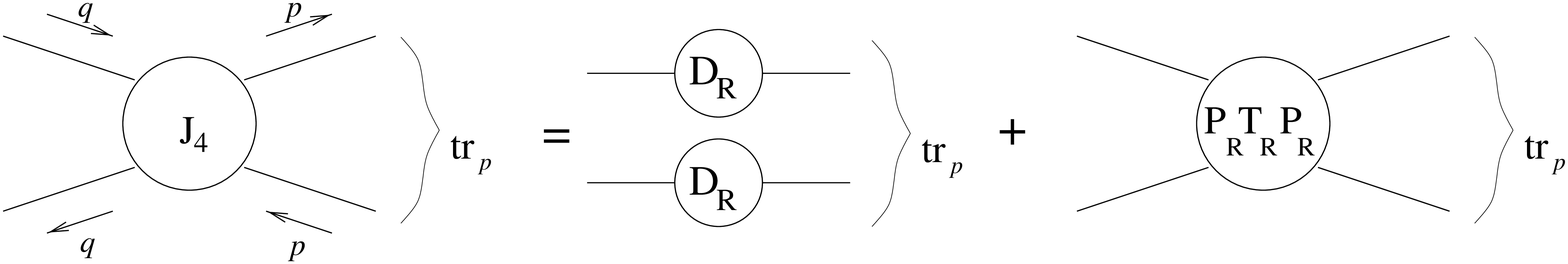}}}
\vskip 1pt
\centerline{Figure 6: Diagrammatic representation (3.17) of $J_{4ik}(q,p)$}

\endinsert
Let us then insert $J_4$ in (3.12d): 
$$\eqalign{
&\sum_j\int{d^nq\over(2\pi)^n}{\rm tr}_q\left\{
\left[{\bf \Gamma}_{\rm R}-{\bf \Gamma}_{\rm R}{\bf P}_{\rm R}
{\bf K}_{\rm R}\right]_{ij}
2^{1-j}J_{4jk}\right\}\cr
=&\sum_j{d^nq\over(2\pi)^n}{\rm tr}_q\left\{
\left[{\bf \Gamma}_{\rm R}-{\bf \Gamma}_{\rm R} 
{\bf P}_{\rm R}{\bf K}_{\rm R}\right]_{ij}(q)\right.\cr
&\left.{\tr}_p\left[{\bf P}_{\rm R}(p)_{jk}(2\pi)^n
\left(\delta(q-p)+\delta_{k,2}\delta
(q+p)\right)+\left[{\bf P}_{\rm R}(q){\bf T}_{\rm R}(q,p)
{\bf P}_{\rm R}(p)\right]_{jk}2^{k-1}
\right]\right\}\cr
=\,&{\rm tr}_p\left\{\left[{\bf \Gamma}_{\rm R}(p){\bf P}_{\rm R}(p)-
({\bf \Gamma}_{\rm R} {\bf P}_{\rm R}{\bf K}_{\rm R}{\bf P}_{\rm R})(p)
+({\bf \Gamma}_{\rm R} {\bf P}_{\rm R}{\bf T}_{\rm R}
{\bf P}_{\rm R})(p)\right.\right.\cr
&~~~~~~~~~~~~~~~~~~~
-\left.\left.({\bf \Gamma}_{\rm R} {\bf P}_{\rm R} 
{\bf K}_{\rm R}{\bf P}_{\rm R}{\bf T}_{\rm R}{\bf P}_{\rm R})(p)
\right]_{ik}2^{k-1}\right\}.
\cr}\eqno(3.18)$$
Using now (3.5d) we see that the last three terms on  the right-hand side
(3.18) cancel, leaving us with the first term which
is obviously finite, being the product of the renormalised
quantities ${\bf \Gamma}_{\rm R}(p)$ and $ {\bf P}_{\rm R} (p)$.

This concludes our discussion of the renormalisation of
the derivatives of the pressure with respect to the masses.
We have shown that in (3.12d) all divergences cancel leaving
us with an ultraviolet finite result.

It remains to make the $1/N$ expansion of the general formula
(3.12d), where only renormalised quantities appear. This is
straightforward but turns out to be a lengthy calculation. We
give some details in the Appendix. As can be seen from there,
starting from (3.12d) the unrenormalised masses and coupling
are never encountered any more and no integration over
negative renormalised squared mass is needed. The differential 
of the pressure with respect to both squared masses is obtained
from (A.5), (A.6), (A.23) as:
$$\eqalign{
        \sum_j dm^2_j{\partial  \over\partial m^2_j} P(T, m_1^2, m_2^2) =
        \sum_j dm^2_j{\partial\over\partial m^2_j}\left\{
        P_0^{(1)}(T, m_1^2) + P_0^{(2)}(T, m_2^2)
        +\,  ^{(0)} \! P^{\rm int}(T, m_1^2, m_2^2)  \right. &\cr
        +{} \left. {\cal O}(1/N)\right\},
~~~~~~~~~~~~~~~~~~~~~~~~~~~~~~~~~~~~~~~~~~~~~~~~~~&
}\eqno(3.19)
$$
$$\eqalign{
P_0^{(1)}(T, m_1^2) &= 4 N \pi \int{d^nq\over(2\pi)^n} n(q)\theta(q^2 - m_1^2),\cr
P_0^{(2)}(T, m_2^2) &= C \pi \int{d^nq\over(2\pi)^n} n(q)\theta(q^2 - m_2^2).}\eqno(3.20)
$$
$P_0^{(1)}(T, m_1^2)$ is the pressure from the free electrons
of renormalised mass $m_1$, $P_0^{(2)}(T, m_2^2)$ similarly
for the photons and $^{(0)}P^{\rm int}(T, m_1^2, m_2^2)$ is the interaction
pressure to order $N^0$ as in eq.\ (2.17). (To be consistent with the
notation of this section we write here $^{(0)}P^{\rm int}$ for the expression
(2.17).)
The mass dependences are indicated explicitly, and we
remind the reader that always the coupling constant
$\tilde g$ is kept fixed.

To summarise: We have calculated the derivatives of the
pressure $\partial P(T, m_1^2, m_2^2)/\partial m^2_j$ to order $N$ and $N^0$ for all masses  
of photons
and electrons satisfying $0\leq m^2_2< 4m^2_1$. Starting from
one point $(m^2_1,m^2_2)$ in this physical region we
can integrate along any path $C_1$ in the $m^2_1-m^2_2$ plane
(figure 7) running to infinite masses, but always staying in the 
physical region.
\topinsert
\centerline{{\epsfxsize=60mm\epsfbox{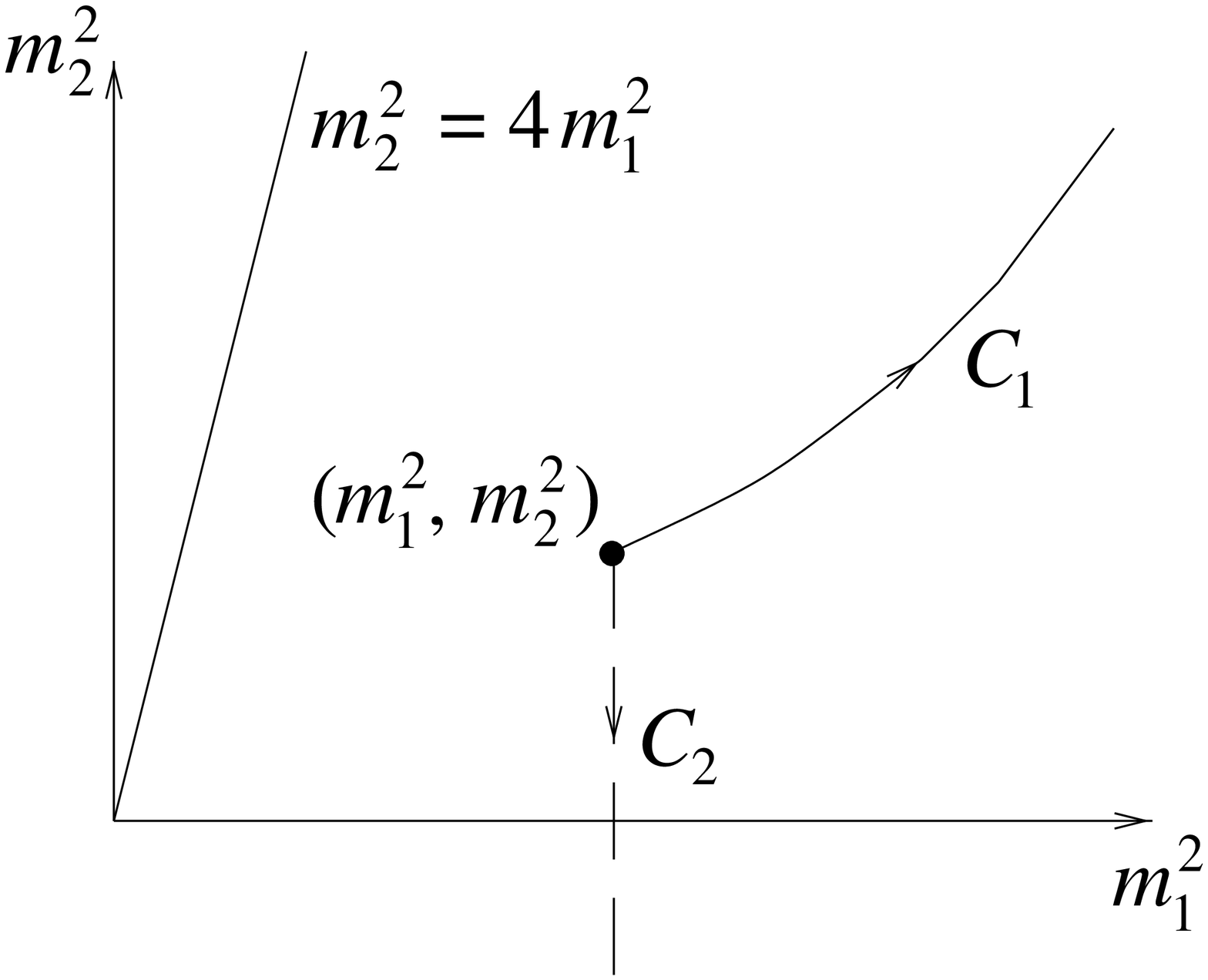}}}
\vskip 1pt
\centerline{Figure 7: Integration path for the mass integration.}
\bigskip
\endinsert
Using then as boundary condition that at infinite masses the pressure vanishes, we
get:
$$\eqalign{
P(T, m^2_1,m^2_2)&=-\int_{C_1}\sum_j\ dm_j^{'2}{\partial\over
\partial m^{'2}_j}P(T,m^{'2}_1,m^{'2}_2) \cr &=
        P_0^{(1)}(T, m_1^2) + P_0^{(2)}(T, m_2^2)
        +\, ^{(0)}\! P^{\rm int}(T, m_1^2, m_2^2)  
        + {\cal O}(1/N)} \eqno(3.21)$$

On the other hand, in the simpler calculation performed
in section 2 we had to integrate along $C_2$, that is
at fixed $m^{'2}_1=m^2_1$ with $m^{'2}_2$ running from
$m^2_2$ to $-\infty$. Clearly, $C_2$ leaves the physical
region. Nevertheless the result obtained in (2.17) is correct,
as we have shown here how it can
be obtained without going to unphysical values for the masses.

Thus the final
result for the pressure to order $N^1$ and $N^0$ is exactly as
described in section 2: To order $N^1$ we get the pressure
of free ``electrons'' and ``positrons'' of
renormalised mass $m_1$, to order $N^0$ we get the free
pressure of ``photons'' with renormalised mass $m_2$ plus
the interaction pressure $^{(0)}P(T)^{int}$ of (2.17).

\bigskip

{\bf 4 Evaluation of the large-$N$ limit}

\def\0{\over}
\def\({\left(} \def\){\right)}
\def\[{\left[} \def\]{\right]}
\def\bq{|{\bf q}|}
\def\qq{{\bf q}^2}
\def\ok{\o_k }

We now turn to the evaluation of the exact result (2.17) which we obtained
in the limit $N\to\infty$. Since (2.17) is UV finite, we can put $\e\to0$.
All formulae in this section thus refer to $n=6$ dimensions.

Equation (2.17) can now be evaluated by a number of nested numerical
integrations which involve the following functions as building blocks.
Firstly, the zero-temperature contributions to the ``photon'' self-energy:
$$
{\rm Re}\; \bar\pi(q^2)=-{1\032\pi^3} \biggl( 
\int_0^1 dx \left[ m_1^2 - q^2 x(1-x) \right]
\log \left|{m_1^2 - q^2 x(1-x)\0m_1^2 - m_2^2 x(1-x)}\right| + {1\06}(q^2-m_2^2) 
\biggr)
\eqno(4.1%repi0
)$$
$$
{\rm Im}\; \bar\pi(q^2)=-{1\032\pi^2} \int_0^1 dx
\left[ q^2 x(1-x) - m_1^2 \right]
\theta(q^2 x(1-x) - m_1^2) = -{1\0192\pi^2} \theta(q^2-4m_1^2)
{(q^2-4m_1^2)^{3/2}\0\sqrt{q^2}}
\eqno(4.2%impi0
)$$
which can be expressed in terms of elementary functions, though we have
done so only for the imaginary part.

Secondly, the thermal contributions. The real part of the
thermal self-energy is given by
$$\eqalign{
{\rm Re}\; \left(
\bar\pi _T (q^0,{\bf q}^2)-\bar\pi(q^2) \right) =
{1\016\pi^3\bq} \int_0^\infty \!dk\; {k^3 n(\ok)\0\ok} \biggl\{&
\[ 1 - \({2\ok q_0-q^2\02k\bq}\)^2\]
\log \left|{\ok q_0- k\bq-\half q^2 \0 \ok q_0+k\bq-\half q^2}\right| \cr
+&\[ 1 - \({2\ok q_0+q^2\02k\bq}\)^2\]
\log \left|{\ok q_0+k\bq+\half q^2 \0 \ok q_0-k\bq+\half q^2}\right| \cr
+&{2q^2\0k\bq} \biggr\} }
\eqno(4.3%repiT
)$$

where $\ok=\sqrt{k^2+m_1^2}$.
The imaginary part is most easily obtained from
$\pi^{12}_T$ via 
$${\rm Im}\;\pi_T=-\half i\e(q_0)(1-e^{\b q_0})\,\pi^{12}_T$$
and using $f(x)f(y)=f(x+y)[1+f(x)+f(y)]$ for $f(x)=1/(e^x-1)$, yielding
$$\eqalign{
{\rm Im}\; \left(
\bar\pi _T (q^0,{\bf q}^2)-\bar\pi(q^2) \right) =
-{\e(q_0)\016\pi^2\bq} &\int_0^\infty \!dk\; {k^3 n(\ok)\0\ok} \biggl\{\cr
&\[ 1 - \({2\ok q_0-q^2\02k\bq}\)^2\] 
\theta\(1-\left|{2\ok q_0-q^2\02k\bq}\right|\)
\e(q_0-\ok) \cr
+&
\[ 1 - \({2\ok q_0+q^2\02k\bq}\)^2\] 
\theta\(1-\left|{2\ok q_0+q^2\02k\bq}\right|\)
\e(q_0+\ok) \biggr\} \cr }
\eqno(4.4%impiT
)$$

In (2.17) we also need $\hat\pi_T$. Its real part is identical to
(4.3), whereas its imaginary part differs from (4.4) by
replacing all the sign functions $\e$ with $1$.

In the limit $\bq\to0$, the integral in (4.4) can be done analytically
with the result
$$
{\rm Im}\;\bar\pi _T (q^0,0) = 
{\rm Im}\;\hat\pi _T (q^0,0) = \( 1+2n({\half q_0}) \)
{\rm Im}\;\bar\pi (q^2=q_0^2)
\eqno(4.5)$$

Another limiting case which can be solved is the high-temperature
limit $|q_0|,\bq,m_1 \ll T$, which gives
$$
\bar\pi _T (q^0,{\bf q}^2)=-{T^2\024\pi} \(1-{q_0^2\0\qq}\)
\[ 1- {q_0\02\bq} \log {q_0+\bq\0q_0-\bq}\] + O(T)
\eqno(4.6%HT
)$$

Remarkably, this is basically the same function that appears in the 
longitudinal
component of the polarization tensor of hot QED and QCD\ref\lebellac,
except that here it comes with a reversed over-all sign.\footnote*{This
abnormal sign in $\phi^3_6$-theories has previously been noted 
in reference\defref\pisnpa{R D Pisarski, Nucl Phys A525 (1991) 175c}.}

As a consequence, the spectrum as read from the
analytic structure of the full thermal ``photon'' propagator
$1/(q^2+m_2^2-g^2 \bar\pi _T (q^0,{\bf q}^2))$
is rather unusual.

In the case of initially
massless ``photons'', the full thermal propagator still has
singularities at the light-cone,
because (4.3) vanishes at $q^2=0$; our massless ``photons''
do not acquire thermal masses. However, there are nontrivial
corrections to the residues of the poles at $q^2=0$ according to
$$\eqalign{
Z(\qq)=&\lim_{q_0\to \bq} (1-g^2 {\partial\0\partial q_0^2} \bar\pi_T)^{-1}\cr
=&\;\(1+{g^2\04\pi^3\qq} \int_0^\infty \!dk\; {k^2 n(\ok)\0\ok}
\[{\ok\02k}\log {\ok+k\0\ok-k}-1\]\)^{-1}  < 1 \cr}
\eqno(4.7)$$
This vanishes as $\bq/T \to 0$ and also as $m_1/T \to 0$.
In the high-temperature limit it is very small for momenta that are not
at least comparable with $T$ in magnitude.
So while no mass gap is
generated, as the temperature increases
thermal effects progressively remove the modes with
larger and larger momenta. In the infinite-temperature limit 
the residue of the pole becomes zero, so that then
there are no propagating plasmons at all.

On the other hand, for non-zero or non-neglible
``photon'' mass, there are always simple poles in the photon propagator.
In this case there are thermal mass corrections, but they are negative,
towards lighter (but nonzero) effective masses. At the same time,
the residues of the corresponding poles are diminished.

At $q_0=0$, $\bar\pi_T$ normally gives the screening mass-squared
for static fields, which in gauge theories is the Debye mass. While screening
corresponds to poles in the propagator for imaginary values of the
spatial momentum, in our model we have a pole at {\it real} spatial momentum
if the temperature is larger than some critical temperature $T_{\rm
crit}$. For $T\gg m_1, m_2$ this pole is located at $\qq=g^2
T^2/(24\pi)$, according to (4.6).  
A similar behaviour has been found in the gravitational polarization
tensor of ultrarelativistic plasmas when evaluated on a flat-space
background \defref\grav{D J Gross, M J Perry and L G Yaffe, Phys Rev
D25 (1982) 330\h
 P S Gribovsky, J F Donoghue and B R Holstein, Ann Phys
(NY) 190 (1989) 149\h
 A Rebhan, Nucl Phys B351 (1991) 706}. There the
value of $\bq$ at the pole at $q_0=0$ is identified with the so-called
Jeans mass characterizing the gravitational instability of the plasma.
In our case, such a pole seems to reflect the fact that the potential
of our model is unbounded from below, so that when the ``photon'' mass
is small enough, thermal fluctuations can lead to a run-away symmetry
breaking without the need of tunneling.

For non-zero $q_0$, there are no poles at space-like momenta, because
for those
(4.6) has a large imaginary part proportional to  $T^2$ corresponding to
Landau damping. So there are no propagating
thermal tachyons.

To summarize, the spectrum of our model in the limit of large $N$
as read from the thermal propagators
is the following. The only thermal corrections occur in the photon spectrum,
which for temperatures sufficiently small compared to the photon mass $m_2$
consist of negative (momentum-dependent) corrections to $m_2$. The latter
are largest at low momenta and tend to zero for very high momenta.
The corresponding dispersion law is depicted
in figure 8a for $g=10$ and $m_1=m_2=m=T < T_{\rm crit}\approx1.236m$.
Up to the critical temperature, the static fields have finite screening
length, which becomes infinite at $T_{\rm crit}$, whereas the
plasma frequency (the long-wavelength limit of the dynamical mass)
remains nonzero. Right at the critical temperature, the spectrum
is thus very similar to that of the transverse vector bosons in the
high-temperature limit of
4-dimensional gauge theories: a vanishing screening mass
together with a nonzero plasmon mass.
For temperatures above $T_{\rm crit}$, there is a pole at
space-like momentum $q_0=0$
and $\qq=m_J^2$ signalling a Jeans-type instability.
As shown in figure 8b for $T=1.5 m>T_{\rm crit}$,
the real part of the inverse photon propagator has
zeros for space-like momenta with $\qq<m_J^2$, which
gives rise to a pole of the propagator where also the
imaginary part vanishes, which is at $q_0=0$.

\topinsert
\centerline{\hfil{\epsfxsize=70mm\epsfbox{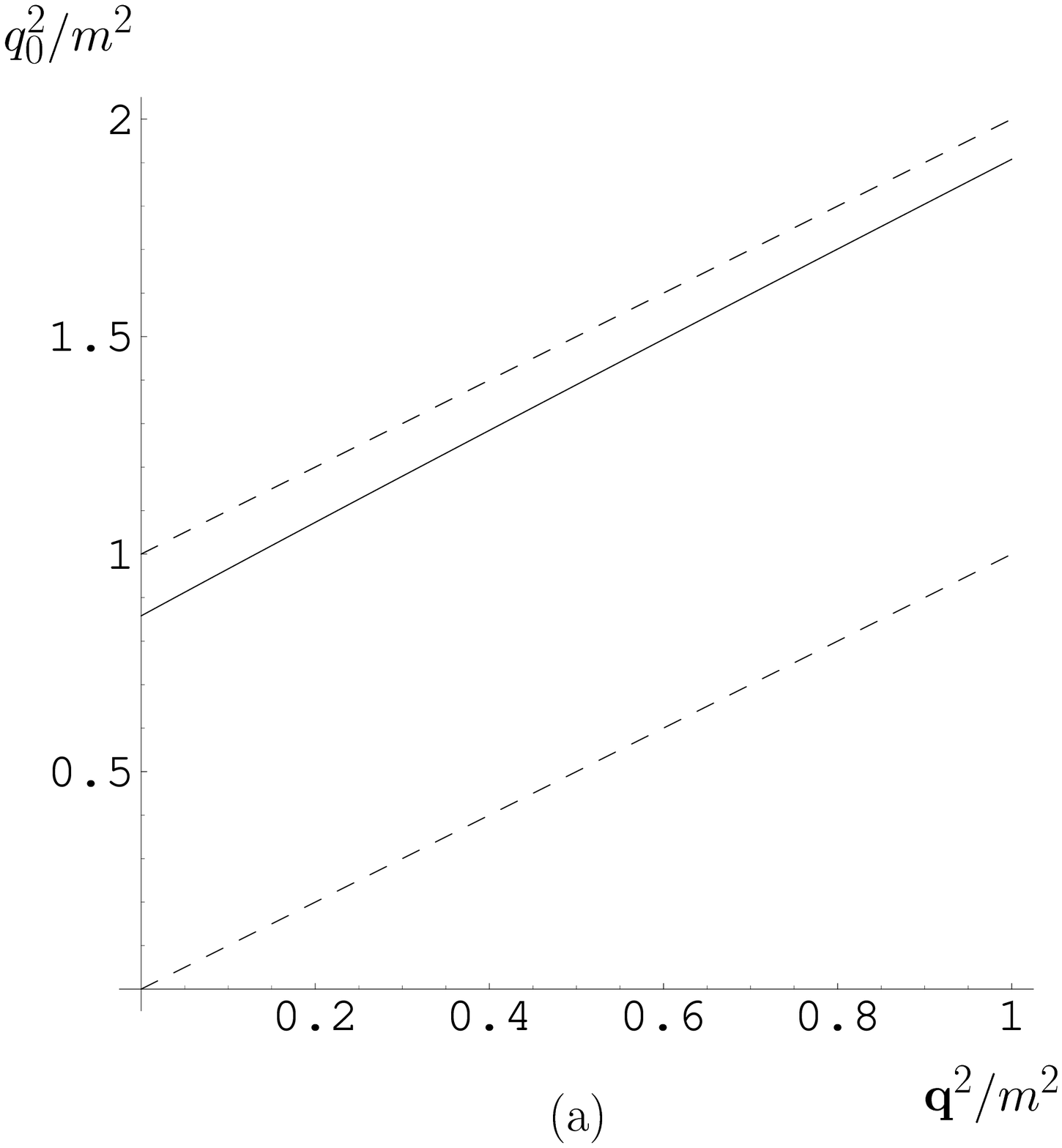}}\hfil{\epsfxsize=70mm\epsfbox{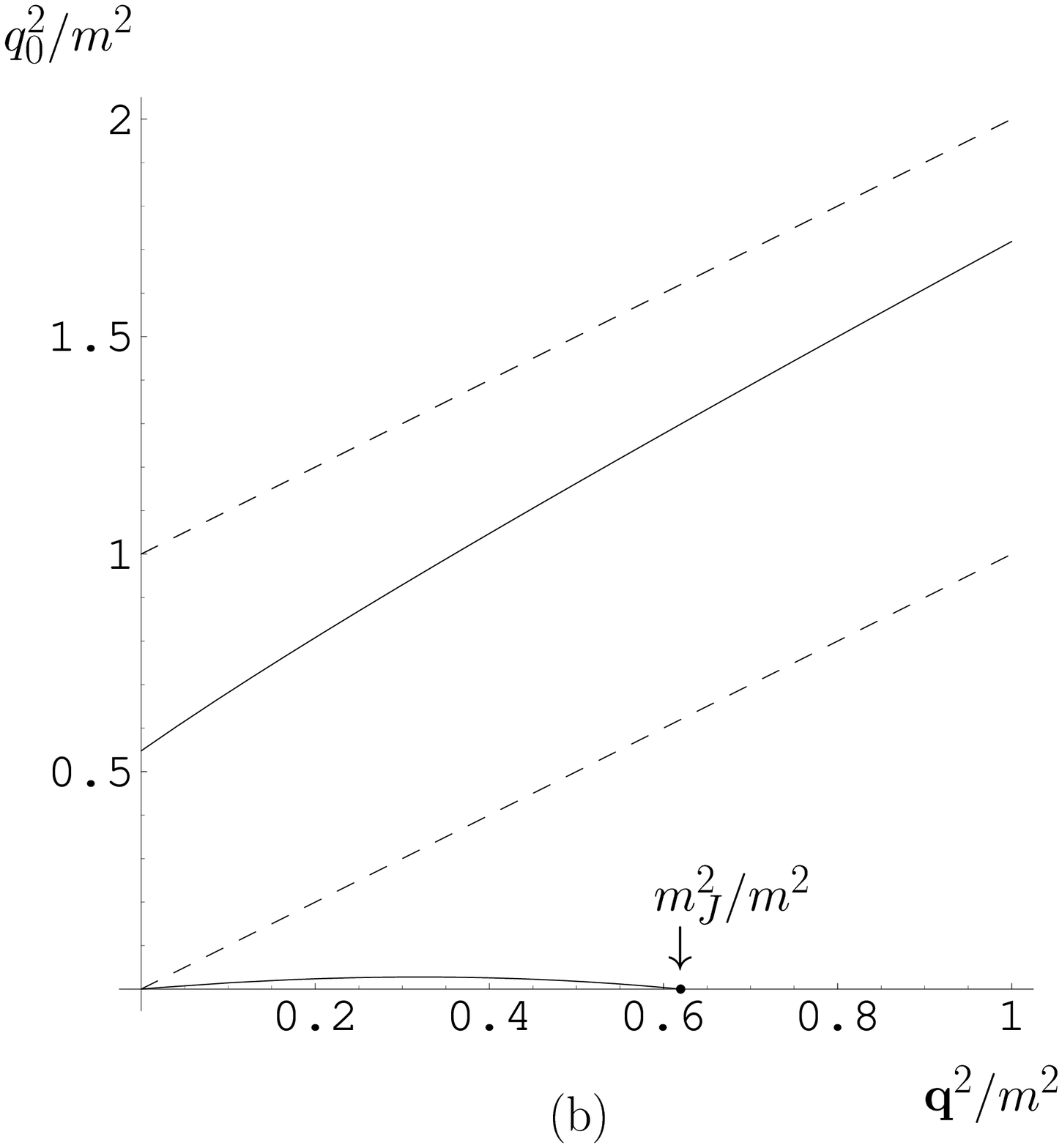}}\hfil\phantom{m}}
\vskip -2pt
\leftskip4mm\rightskip4mm
Figure 8: The location of poles in the ``photon'' propagator
for $g=10$ at $T=m=m_1=m_2$ (a), which is below  the critical temperature, and 
at $T=1.5m$ (b), which is above it. The two
dashed lines mark the light cone and the zero-temperature mass
hyperboloid $q_0^2=\qq+m_2^2$. The  line between them
is the location of the poles in the thermal propagator. In (b)
the second full line below the light cone marks zeros of the
real part of the inverse propagator, which correspond to
poles of the propagator only at its intersections with $q_0=0$
because of large Landau damping for $q_0>0$ and $q^2<0$.
\endinsert

The presence of a pole at $q_0=0$ and $\bq=m_J>0$ causes the
interaction pressure
$$\eqalign{
P(T)^{\rm int}
&=-\half C  \int {d^{6}q\over (2\pi)^{6}}
\Bigl[ 2n(q) \arg {g^2\bar\pi _T (q^0,{\bf q}^2,m_2^{2})+m_2^2-q^2
\over m_2^{2}-q^2} \cr
&\qquad +\arg {g^2\bar\pi _T (q^0,{\bf q}^2,m_2^{2})+m_2^2-q^2
\over g^2\bar\pi (q^2,m_2^{2})+m_2^2-q^2}
+g^2 \hbox{ Im }{\hat\pi _T (q^0,{\bf q}^2,m_2^{2})\over
q^2-m_2^{2}-g^2\bar\pi (q^2,m_2^{2})} \Bigr] }
\eqno(2.17)
$$
to become IR singular for all $T>T_{\rm crit}$. 
In the first term, for $|q_0|$ smaller
than some finite number, the real
part of $g^2\bar\pi _T+m_2^2-q^2$ changes sign from positive to negative
as $\bq$ is changed from some large value to a sufficiently small one
(see figure 8b).
The imaginary part on the other hand is negative throughout except
for $|q_0|=0$, where it vanishes. The argument in the first term
therefore approaches
a step function $-\pi\theta(m_J-\bq)$ as $|q_0|\to0$. This causes
the first term in (2.17) to diverge logarithmically in the IR
when $T>T_{\rm crit}$.

For finite photon mass and $T\le T_{\rm crit}$, the
pole at $q_0=0$ and $\bq=m_J>0$ is absent and 
(2.17') appears to be well-defined.
%$m_1=m_2=m$ and $g=1$,
%the critical temperature above which (2.17) ceases to exist
%turns out to be $T_{\rm crit}\approx 9.13 m$.
%Thus there is a nonzero
%plasma frequency, which at the critical temperature in our
%example is $m_{pl.}\approx 0.82 m$ for $m_1=m_2=m$ and $g=1$.

Turning finally to the numerical evaluation of (2.17),
we can distinguish three different regions
of integration depending on the appearance of imaginary
parts.

The inverse photon propagator $g^2\bar\pi _T+m_2^2-q^2$
has imaginary parts only for $q^2>4 m_1^2$
(pair creation) and for $q^2<0$ (Landau damping). The quasi-particles
described by the time-like poles of the photon propagator are therefore
undamped and stable, provided $m_2 < 2m_1$. 
We denote their position by $\o_T(\qq)$.
For $|q_0| < \o_T(\qq)$ 
the real part of $g^2\bar\pi _T+m_2^2-q^2$ is positive 
and for $|q_0| > \o_T(\qq)$ it is negative.

Correspondingly, we have:

$I$ --- spacelike momenta, $q^2<0$: $\bar\pi$ is real in
this region, but $\bar\pi _T$ has imaginary parts corresponding
to Landau damping, so that
$$\eqalign{
P(T)^{\rm int}_I
=-C  \int {d^{5}{\bf q}\over (2\pi)^{6}} \int_0^{\bq} d q_0&\Bigl\{
 (2n(q_0)+1) \arg [g^2\bar\pi _T (q^0,{\bf q}^2,m_2^{2})+m_2^2-q^2] \cr
& + g^2 { \hbox{ Im }\hat\pi _T (q^0,{\bf q}^2,m_2^{2}) \over
q^2-m_2^{2}-g^2\bar\pi (q^2,m_2^{2})} \Bigr\} }
\eqno(4.8a)
$$

$II$ --- timelike momenta below threshold, $0<q^2<4 m_1^2$:
the first two terms in (2.17) contribute only for $\sqrt{\qq+m_2^2}\ge
|q_0|\ge \o_T(\qq)$, whereas the last term has a pole in this
range with unit residue thanks to our on-shell renormalization scheme.
The integration over $q_0$ can be carried out with the result
$$\eqalign{
P(T)^{\rm int}_{II}
=-C  \int {d^{5}{\bf q}\over (2\pi)^{6}} (-\pi)&\biggl\{
2 \log{e^{\sqrt{\qq+m_2^2}}-1\over e^{\o_T(\qq)}-1}+
\o_T(\qq)-\sqrt{\qq+m_2^2}\cr
&+g^2{\hbox{ Re }\bar\pi_T(q^0=\sqrt{\qq+m_2^2},{\bf q}^2,m_2^{2})\over 
2\sqrt{\qq+m_2^2}} \biggr\}}
\eqno(4.8b)
$$

$III$ --- timelike momenta above threshold, $q^2 > 4 m_1^2$:
$$\eqalign{
P(T)^{\rm int}_{III}
=-C & \int {d^{5}{\bf q}\over (2\pi)^{6}} \int_{\sqrt{\qq+4 m_1^2}}^\infty d q_0
\biggl\{
 2n(q_0) \Bigl(\arg \[g^2\bar\pi _T (q^0,{\bf q}^2,m_2^{2})+m_2^2-q^2\]+\pi\Bigr)\cr
&+\arg \[g^2\bar\pi _T (q^0,{\bf q}^2,m_2^{2})+m_2^2-q^2\]-
\arg \[g^2\bar\pi(q^2,m_2^{2})+m_2^2-q^2\]  \cr
&+g^2 \hbox{ Im }{\hat\pi _T (q^0,{\bf q}^2,m_2^{2})\over
q^2-m_2^{2}-g^2\bar\pi (q^2,m_2^{2})} \biggr\}
}
\eqno(4.8c)
$$

\topinsert
\centerline{{\epsfxsize=\hsize\epsfbox{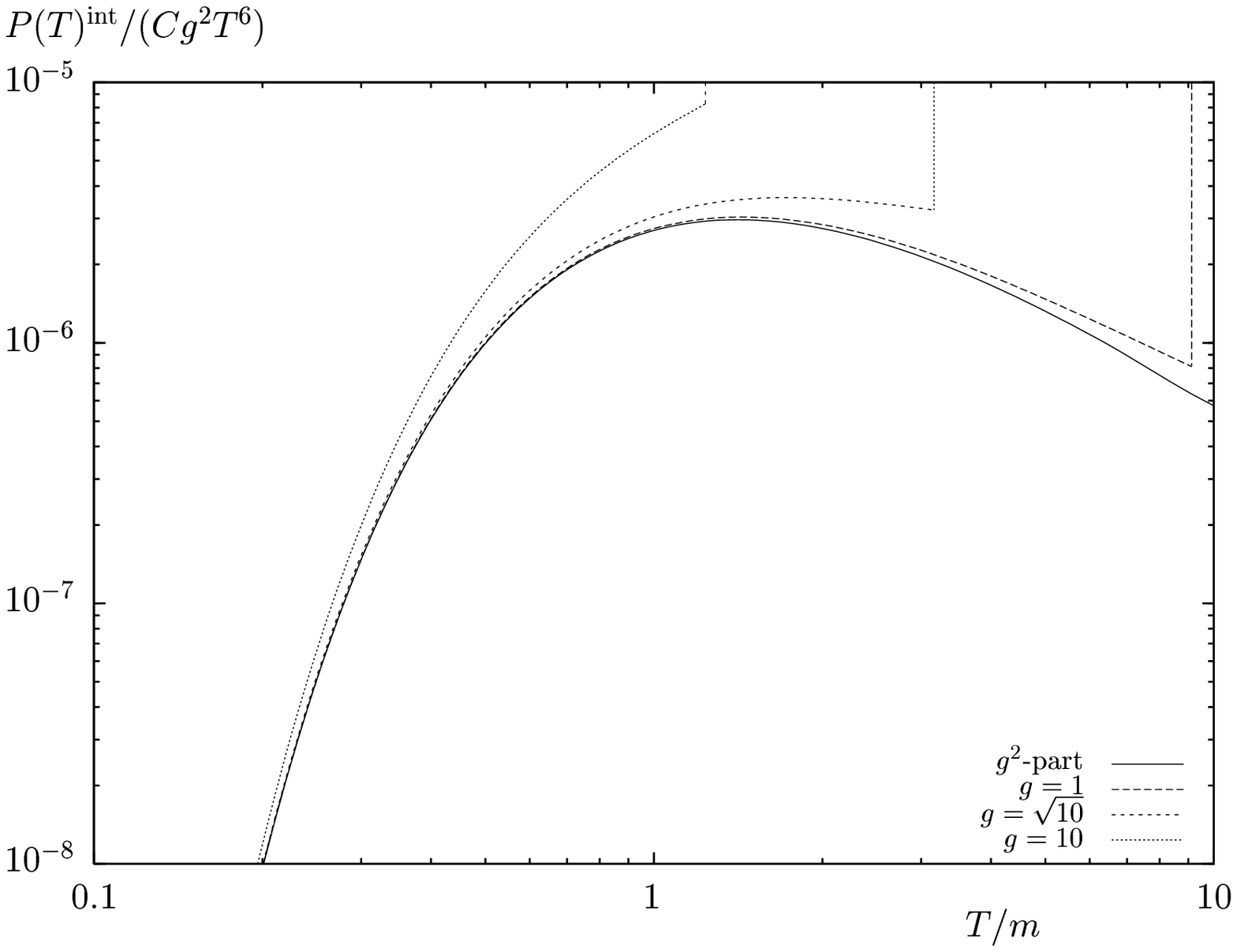}}}
\vskip -5pt
\leftskip4mm\rightskip4mm
{Figure 9: Numerical evaluation of the interaction
pressure (2.17) for three
values of $g$}
{in comparison with the leading perturbative contribution.}
\endinsert

The numerical evaluation of these expressions is quite challenging
because of large cancellations among the individual contributions so that
rather high working precision is needed. Our results for the nonperturbative
interaction pressure are given in figure 9
for $m=m_1=m_2$ and three different values of the coupling, $g=1$, $\sqrt{10}$,
and 10, and these are compared with the strictly perturbative
contribution proportional to $g^2$, which also has to be computed
numerically because we do not restrict ourselves to the high-temperature
limit and so the full momentum dependence of the thermal self-energy
enters there too.

For temperatures $T\ll m$ there is rather little difference between
the perturbative and the nonperturbative results. As the temperature
increases, the latter grow bigger than the former until they abruptly
end in a singularity at the critical temperature.
For sufficiently small coupling and for $m_1\sim m_2$, 
the critical temperature  above which (2.17) 
ceases to exist can be estimated from the high-temperature
expansion of the thermal photon propagator following from (4.6):
$T_{\rm crit} \approx  \sqrt{24\pi} m_2 /g \approx 9 m_2 /g$. 
The actual values for $g=1,
\sqrt{10},10$ are 
$T_{\rm crit} \approx 9.133,\, 3.164,\, 1.236$ times $m$, respectively. 
At these temperatures the thermal pressure ceases to exist, because
there is sort of a phase transition to a run-away and therefore
inexistent broken phase.

At exactly $T=T_{\rm crit}$, we have a situation which is closest
to gauge theories in 4 dimensions, because there $m_J=0$,
corresponding
to a vanishing screening mass as is the case for
the magnetostatic modes in perturbation theory.
The interaction pressure is still
well-defined and is given by the end-points of the various
curves in figure 9. A conspicuous difference from the results
of 4-dimensional gauge theories is that the interaction pressure
is positive, which is related to the abnormal sign of
all thermal mass corrections in our model.  

\bigskip\goodbreak
{\bf 5 Conclusions}

We have pursued two purposes in our study of our six-dimensional
scalar model. Firstly, we have investigated how precisely the nonperturbative
formula for the thermal pressure proposed in reference~{\first} is rendered
finite by standard zero-temperature renormalisation.

The renormalisation of the derivatives of the pressure with respect to
the masses leads us to a problem of overlapping divergences which we
solved in a manner analogous to Dyson's method for QED,
introducing a certain 2--2 scattering kernel.  We derived the
renormalised Dyson-type equations, which turned out to be essential for
our discussion of renormalisability. Finally, we had to invoke Wilson's
operator product expansion. We think that we can draw the lesson from
this that in more complicated theories like QCD things will not be
simpler and one will again have to deal with overlapping divergence
problems. On the other hand, having an expression for the pressure in
terms of renormalised Green's functions as given for our model in
(3.12d) and knowing the essential equations these Green's functions
must satisfy in order to have ultraviolet finiteness, may help to
devise consistent approximation schemes leading to finite results in
all orders. With these methods we should also be able 
to study explicitly the effects occuring when a particle---in our case
the ``photon''---becomes unstable. In the Green's functions for
$T$=0 this amounts simply to a pole moving from the first to the
second Riemann sheet and thus we expect that our result for the pressure,
which is expressed in terms of these functions, should not change 
drastically.

Besides these general aspects, we have investigated the large $N$ limit
of our model which diagrammatically is similar to QED in the limit
of large flavour numbers. In this limit the leading contribution
to the interaction pressure comes from a ring resummation of
the photon polarization function, while the electron lines remain
undressed. When $N$ is not large, keeping only this
contribution corresponds to what is
known as random-phase approximation (RPA)\defref\BP{D Bohm and D Pines, 
Phys Rev 92 (1953) 609} 
in many-body physics. In contrast to the simpler ring resummation
of the Debye screening mass\defref\GMB{M Gell-Mann and K A Brueckner,
Phys Rev 106 (1957) 364}, 
one has to deal with a resummation of
a momentum-dependent quantity. In practice, however, one usually
aims at an (improved) perturbative scheme and uses this resummation
only as far as needed to extract the next-to-leading order term
in the interaction pressure, which because of the infrared
singularities in the usual series is nonanalytic\defref\AP{I A
Akhiezer and S V Peletminskii, Sov Phys JETP 11 (1960) 1316}
\defref\FMcL{B A Freedman and L McLerran, Phys Rev D16 (1977) 1130,
1147, 1169}  in $e^2$.
Here we have found that the RPA can be interpreted as the
leading term in a large-$N$-expansion and we have retained
the full nonperturbative information that it incorporates.

In our six-dimensional scalar model we have in fact encountered
rather drastic resummation effects, because this model has a critical
temperature above which an instability similar to the gravitational
Jeans instability occurs. So despite the diagrammatic similarity
with QED, this theory is rather different from it.
But below the critical temperature we were able to obtain
a nonperturbative expression for the interaction pressure, and
evaluate it  numerically. 
Right at the critical temperature,
where the nonperturbative interaction pressure is still well-defined,
the spectrum of our model is even rather similar to that of
perturbative four-dimensional
gauge theories in that it has a vanishing screening mass
like the magnetostatic modes.

The computation of the nonperturbative interaction pressure
in more realistic theories such as ordinary QED 
in the limit of large flavour numbers
would be technically not too different from what we have
done here. Similar simplifications seem to be of interest
even in QCD in the small-$N_c$ large-$N_f$ limit\defref\BH{S J Brodsky
and P Huet, Phys Lett B417 (1998) 145}.
We plan to investigate those theories along
the above lines in a separate work.

\bigskip
\bigskip{\eightit
This research is supported in part by the EU Programme
``Training and Mobility of Researchers", Networks
``Hadronic Physics with High Energy Electromagnetic Probes"
(contract FMRX-CT96-0008) and
``Quantum Chromodynamics and the Deep Structure of
Elementary Particles'' (contract FMRX-CT98-0194),
by the British Council and the German Academic Exchange Service
(ARC projects 577 and 313),
by the Jubil\"aumsfonds der
\"Osterreichischen Nationalbank (project no 5986), by
the Leverhulme Trust and by PPARC}
\bigskip\goodbreak
\bigskip\bigskip
\def\0{\over}
\def\({\left(} \def\){\right)}
\def\[{\left[} \def\]{\right]}
\def\bq{|{\bf q}|}
\def\qq{{\bf q}^2}
\def\ok{\o_k }

{\bf Appendix}

In this appendix we show how the $1/N$ expansion of the
general formula (3.12d) leads us to the results of section
2. In the first step we perform all the traces implicit in
(3.12d). For this we define:
$$
\hat K_{jl\rm R}(q,p)=i2^{2-j-l}({\tr}_q
{\tr}_p)(K_{jl\R}(q,p)),\eqno(A.1)$$

\topinsert
\centerline{{\epsfxsize=60mm\epsfbox{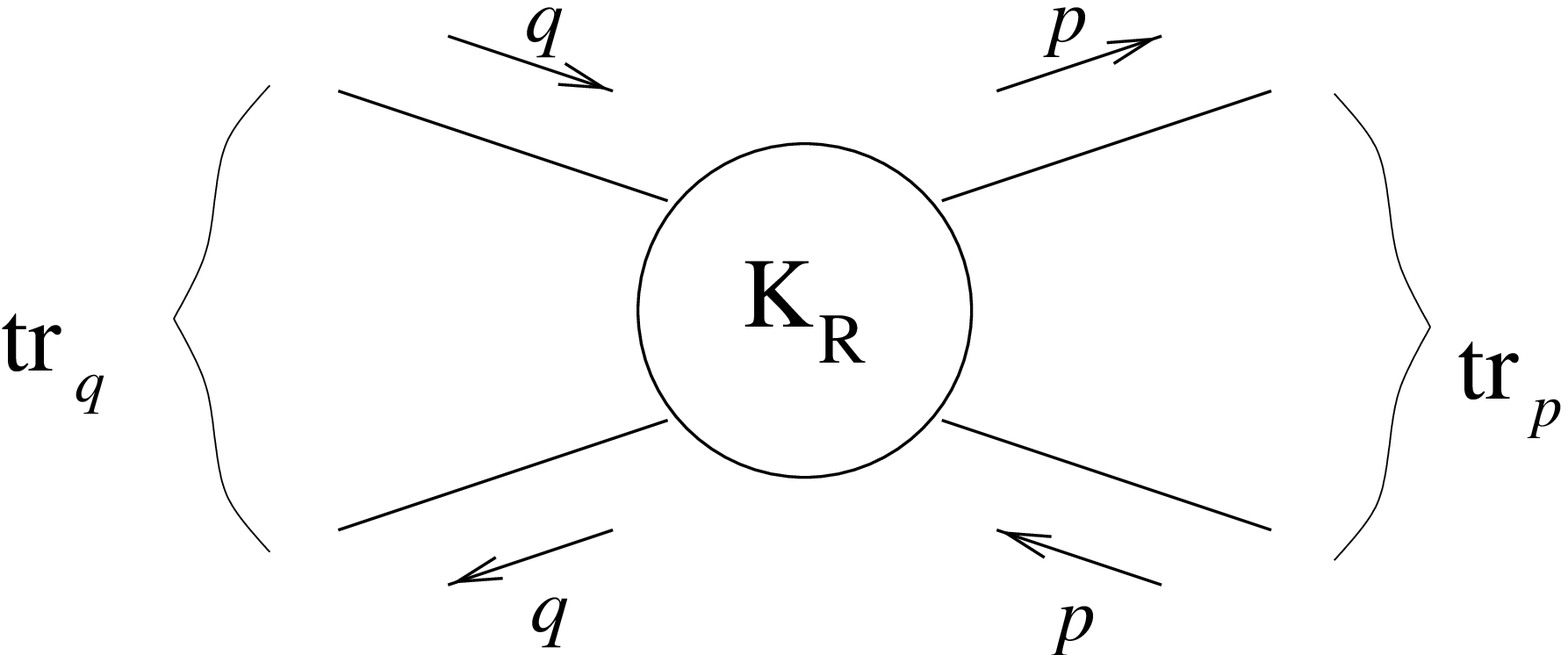}}}
\vskip 1pt
\centerline{Figure 10: Diagrammatic picture for the
expression (A.1)}
\endinsert

where tr$_q$ (tr$_p$) means the trace over the
internal indices of the lines where the momentum $q\ (p)$
flows (figure 10). We also define  the differential forms:
$$
\omega_j(p)=\sum_ldm^2_l\ \hat\Gamma_{lj{\rm R}}(p^2),\eqno(A.2)$$
which can be expressed through the self-energy functions
using the renormalised version of~(3.9c):
$$\eqalign{
&\omega_j(p)=\sum_l\Big\{-dm^2_l{\partial\over\partial  
m^2_l}\left[p^2-m^2_j-\bar\Pi_j(p^2)\right]
+\left[p^2-m^2_j-\bar\Pi_j(p^2)\right]dm^2_l
{\partial\over\partial m^2_l}\log {Z_j\over\tilde Z_j}\Big\}
\cr}\eqno(A.3)$$
Here $\tilde Z_j$ are the wave function renormalisation constants
for the theory with the coupling $\tilde\lambda$, but masses
$\mu_j$ (compare the paragraph following (3.11e)). We find then
for the pressure from (3.12d)
$$\eqalign{
\sum_jdm^2_j{\partial\over\partial m^2_j}P(T)
=&-\int{d^nq\over(2\pi)^n}\Big\{\omega_1(q)2N\hat\Delta^{11}_{1T{\rm R}}(q)
+\omega_2(q){C\over2}\hat\Delta^{11}_{2T{\rm R}}(q)\cr
&-i\int{d^np\over(2\pi)^n}\sum_{j,l}\omega_j(q)\left[q^2-m^2_j-\bar
\Pi_j(q^2)\right]^{-2}\hat K_{jl{\rm R}}
(q,p)\hat\Delta^{11}_{lT{\rm R}}(p)\Big\}\cr}\eqno(A.4)$$
This is the starting point for the $1/N$ expansion. For any
(scalar) quantity $F$ we write $F=\sum_r\ {}^{(r)}F$ with $^{(r)}F$
the term of order $(1/N)^r$. We get then from (A.4)
for the term of order $N$
$$\eqalign{
\sum_jdm^2_j{\partial\over\partial m^2_j}\ {}^{(-1)}P(T)
&=-2N\int{d^nq\over(2\pi)^n}\ {}^{(0)}\omega_1(q)\cdot\ {}^{(0)}\hat
\Delta_{1T{\rm R}}^{11}(q)\cr
&=-2Ndm^2_1\int{d^nq\over(2\pi)^n}n(q)\,2\pi\delta(q^2-m^2_1).\cr}
\eqno(A.5)$$
Therefore, to order $N$ we get the pressure from the free
electrons and positrons of mass $m_1$, that is $P_0^{(1)}(T, m_1^2)$ of (2.10):
$$^{(-1)}P(T)=   P_0^{(1)}(T, m_1^2).\eqno(A.6)$$

To order $N^0$ we get from (A.4) after some work, and using the fact that
the leading contributions to ${\bf K}$ are single-particle exchanges:
$$\eqalign{
\sum_jdm^2_j{\partial  \over\partial m^2_j}\ &{}^{(0)}P(T)=-\int
{d^nq\over(2\pi)^n}\Big\{
2N\Bigg[ -\sum_jdm^2_j{\partial\over\partial m^2_j}
\ {}^{(1)}\Pi_1(m^2_1) \cr
&~~~~~~~~~~~~~~~~~~~~~~~~~~~~~~~~~~~~~~~~~~+dm^2_1\ {}^{(1)}\Pi_1'(m^2_1)\Bigg]
\ {}^{(0)}\hat\Delta^{11}_{1T{\rm R}}(q)\cr
&+2Ndm^2_1{}~{}^{(1)}\hat\Delta^{11}_{1T{\rm R}}(q)
-\half C\ {}^{(0)}Z_2\left[\sum_jdm^2_j{\partial
\over\partial m^2_j}\left (^{(0)}\Pi_2(m^2_2)-m^2_2\right)\right]
  {}^{(0)}\hat\Delta^{11}_{2T{\rm R}}(q)\Big\},~~~~~~~~~~~~~\cr}\eqno(A.7)
$$
where, with the notation of section 2,
$$^{(1)}\Pi_1(q^2)={ig^2\over N}C\int{d^nk\over(2\pi)^n}\left
[(q-k)^2-m^2_1\right]^{-1}\left[k^2-m^2_2-g^2\bar
\pi\left(k^2,m^2_2\right)\right]^{-1},\eqno(A.8)$$
$$^{(0)}\Pi_2(q^2)=g^2_0\pi(q^2),\eqno(A.9)$$
$$^{(0)}Z_2=\left[1-g^2_0\pi'(m^2_2)\right]^{-1}.\eqno(A.10)$$
The unrenormalised quantities $\Pi_{1,2},\ Z_2,\ g_0$
occurring in (A.7) are to be understood as functions
of the renormalised parameters. Of course the individual terms
in (A.7) contain divergences, which must cancel in the sum
as we know from sections 2,3 and as we will again see explicitly
below.

Now we use the general sum rule
$$
\int dq^0\left[\hat\Delta_{jT{\rm R}}^{11}(q)-\hat\Delta^{12}_{jT{\rm R}}
(q)\right]=0\eqno(A.11)$$
which we already mentioned in connection with (1.1d),
to express the integrals over $\hat\Delta^{11}$ in (A.7)
by integrals over $\hat\Delta^{12}$ which in turn is related to
the imaginary part of the thermal propagators. In this way
we obtain
$$
\sum_jdm^2_j{\partial\over\partial m^2_j}\ {}^{(0)}P(T)=\eta_1+\eta_2+
\eta_3,\eqno(A.12)$$
where the differential forms $\eta_{1,2,3}$ are given by
$$\eqalignno{
\eta_1&=2N\int{d^nq\over(2\pi)^n}
\left[\sum_jdm^2_j{\partial\over\partial m^2_j}\ {}^{(1)}\Pi_1(k^2)\right
]_{k^2=m^2_1}
n(q)2\pi\delta(q^2-m^2_1)&(A.13)\cr 
&~~~~~~\cr
\eta_2&=2Ndm^2_1\int{d^nq\over(2\pi)^n}
{\rm Im}\left\{{1+2n(q)\over(q^2-m^2_1)^2}\ {}^{(1)}\bar\Pi_{1T}(q)-
{1\over(q^2-m^2_1)^2}\ {}^{(1)}\bar\Pi_1(q^2)\right\}&(A.14)\cr
&~~~~~~\cr
\eta_3&={1\over2}C
\left[dm^2_2-g^2{\partial\pi(m^2_2)
\over\partial m^2_1}dm_1^2\right
]\int{d^nq\over(2\pi)^n}
{\rm Im}\Big[{1+2n(q)\over q^2-m^2_2-g^2\bar\pi_T(q,m^2_2)}\cr
&~~~~~~~~~~~~~~~~~~~~~~~~~~~~~~~~~~~~~~~~~~~~~~~~~~~~~~~~~~~~~~~
-{1\over q^2-m^2_2-g^2\bar\pi(q^2,m^2_2)}\Big]&(A.15)\cr}
$$
with
$$
{}^{(1)}\bar\Pi_1(q^2)={}^{(1)}\Pi_1(q^2)-{}^{(1)}\Pi_1(m^2_1)
-(q^2-m^2_1)^{(1)}\Pi_1'(m^2_1)\eqno(A.16)$$
$$\eqalignno{
&{}^{(1)}\bar\Pi_{1T}(q)={}^{(1)}\bar\Pi_1(q^2)\cr
&~~~~~~~+{g^2C\over N}\int
{d^nk\over(2\pi)^n}\Big\{
\left[{i\over(q-k)^2-m^2_1}+n(q-k)2\pi\delta((q-k)^2-m^2_1)
\right]\left[k^2-m^2_2-g^2\bar\pi_T(k,m^2_2)\right]^{-1}\cr
&~~~~~~~-{i\over(q-k)^2-m^2_1}\left[k^2-m^2_2-g^2\bar\pi(k^2,m^2_2)\right]
^{-1}\cr
&-i\Big[-{in(k)\over(q-k)^2-m^2_1}+2\pi\delta\left((q-k)
^2-m^2_1\right)
\Big[n(k)\left(\theta(q^0)\theta(k^0-q^0)+\theta(-q^0)
\theta(q^0-k^0)\right)\cr
&~~~~~~~~~~~~~~
 +n(q-k)\left(\theta(q^0)\theta(-k^0)+\theta(-q^0)\theta(k^0)\right)
\Big]\Big]2 {\,\Im}\,\left[k^2-m^2_2-g^2\bar\pi_T(k,m^2_2)
\right]^{-1}\Big\}&(A.17)\cr}
$$
Here ${}^{(1)}\bar\Pi_1(q^2)$
is the renormalised electron self-energy function to order $1/N$ and 
${}^{(1)}\bar\Pi_{1T}(q)$ is the corresponding thermal function.

The forms $\eta_1$ and $\eta_3$ have already a simple structure.
The form $\eta_2$ which arises from the $1/N$ term
${}^{(1)}\hat\Delta^{11}_{1T{\rm R}}$ of the
electron propagator 
in (A.7) is more difficult to handle. The strategy
is to insert (A.16) and (A.17) in (A.14), which leads to 
integrals over $q$ and $k$, and then to perform first
the $q$-integration. In this way we get after some nontrivial
calculations:
$$\eqalign{
&\eta_2=2N\ dm^2_1\left[
{}^{(1)}\Pi_1(m^2_1){\partial\over\partial m^2_1}
+{}^{(1)}\Pi_1'(m^2_1)\right]
\int{d^nk\over(2\pi)^n}n(k)2\pi\delta(k^2-m^2_1)\cr
&+{C\over2}g^2\ dm^2_1\int{d^nk\over(2\pi)^n}\ {\rm Im}\Big\{
(1+2n(k))\left[k^2-m^2_2-g^2\bar\pi_T(k,m^2_2)\right]^{-1}
{\partial\over\partial m^2_1}\pi_T(k)\cr
&-\left[k^2-m^2_2-g^2\bar\pi(k^2,m^2_2)\right]^{-1}
{\partial\over\partial m^2_1}\pi(k^2)\Big\}.\cr}
\eqno(A.18)$$
Putting now everything together, we arrive at:
$$\eqalign{
\sum_j dm^2_j {\partial \over \partial m_j^2} {}^{(0)}P(T)
=& -\sum_j dm^2_j {\partial \over \partial m_j^2}
\left[ ^{(1)}\Pi_1(m_1^2){\partial P_0^{(1)}(T, m_1^2)\over \partial m_1^2}\right]\cr
&{}-dm_1^2 {1\over 2} g^2 C \Im \int{d^n q\over (2\pi)^n}
\left\{\left[q^2 - m_2^2 - g^2 \bar{\pi}(q^2, m_2^2)\right]^{-1}
{\partial \over \partial m_1^2} \left(\pi(q^2) - \pi(m_2^2)\right)
\right.\cr
&\left. {}- [1 + 2 n(q)] 
\left[q^2 - m_2^2 - g^2 \bar{\pi}_T(q, m_2^2)\right]^{-1}
{\partial \over \partial m_1^2} \left(\pi_T(q) - \pi(m_2^2)\right)
\right\}\cr
&{}-dm_2^2 {1\over 2}  C \Im \int{d^n q\over (2\pi)^n}
\left\{\left[q^2 - m_2^2 - g^2 \bar{\pi}(q^2, m_2^2)\right]^{-1} \right.\cr
&{} -\left. [1 + 2 n(q)] 
\left[q^2 - m_2^2 - g^2 \bar{\pi}_T(q, m_2^2)\right]^{-1}
\right\}
}\eqno(A.18a)
$$
In the first term on the right hand side of (A.18a) we can use:
$$\eqalign{
 ^{(1)}\Pi_1(m_1^2){\partial P_0^{(1)}(T, m_1^2)\over \partial m_1^2} = 
{1\over 2} \left[ {\partial P_0^{(1)}(T, m_1^2)\over \partial m_1^2} + c.c. \right]=
~~~~~~~~~~~~~~~~~~~~~~~~~~~~~~~~~~~~~~~~~~~~~~~~~~~~~~~~~~~~~~~~~~~~~~~~~~&\cr
2 C g^2 \Im \left\{ \left. \int{d^n q\over (2\pi)^n}
{1\over (k - q)^2 - m_1^2} {1\over q^2 - m_2^2 - g^2 \bar{\pi}(q^2, m_2^2)}
\right|_{k^2 = m_1^2}
\int{d^n k\over (2\pi)^n} n(k) 2\pi\delta(k^2 - m_1^2)
\right\} ~~~~~~~~~~~~~~~~~~~~~~~~~~~~&
\cr =  2 C g^2 \Im \left\{ \int{d^n q\over (2\pi)^n}
{1\over q^2 - m_2^2 - g^2 \bar{\pi}(q^2, m_2^2)}
\int{d^n k\over (2\pi)^n} n(k) 2\pi\delta(k^2 - m_1^2)
{1\over (k - q)^2 - m_1^2}\right\}~~~~~~~~~~~~~~~~~~~~~~~~~~~~~~~~~~~~ &\cr
={1\over 2} C g^2 \Im \int{d^n q\over (2\pi)^n}
{\hat{\pi}_T(q)\over q^2 - m_2^2 - g^2 \bar{\pi}(q^2, m_2^2)}
~~~~~~~~~~~~~~~~~~~~~~~~~~~~~~~~~~~~~~~~~~~~~~~~~~~~~~~~~~~~~~~~~
&}\eqno(A.18b)
$$
This is how the function $\hat{\pi}_T(q)$ (2.11d) appears in the present way
of calculating the pressure.

The last relation we need is the one between the coupling constants
$g$ and $\tilde g=\tilde\lambda \sqrt N$ (compare the paragraph
before (3.11a).) To order $N^0$ we have
$$
g^2=\tilde g^2\left\{1+\tilde g^2\left[\tilde\pi'(\mu^2_2)-\pi'(m^2_2)
\right]\right\}^{-1},\eqno(A.19)$$
where $\tilde\pi(q^2)$ is defined as $\pi(q^2)$ in (2.12a), but with
mass $\mu_1$ instead of $m_1$. From (A.19)
we find for the derivatives of
$g$ with respect to $m^2_{1}$ and $m^2_{2}$ keeping $\tilde g$ fixed:
$$
{\partial g^2\over\partial m^2_j}=(g^2)^2
{\partial\over\partial m^2_j}\left[
\pi'(m^2_2)\right],\quad(j=1,2).\eqno(A.20)$$
From this we derive easily
$$\eqalign{
&{\partial\over\partial m^2_1}\left[g^2\bar\pi_T(k,m^2_2)\right]
=g^2{\partial\over\partial m^2_1}\left[\pi_T(k)-\pi(m^2_2)\right]\cr
&-\left[k^2-m^2_2-g^2\bar\pi_T(k,m^2_2)\right]
g^2{\partial\over\partial m^2_1}\pi'(m^2_2),\cr}
\eqno(A.21)$$
$$
{\partial\over\partial m^2_2}\left[g^2\bar\pi_T(k,m^2_2)\right]
=-g^2\left[k^2-m^2_2-g^2\bar\pi_T(k,m^2_2)\right]
\pi''(m^2_2)\eqno(A.22)$$
Collecting everything together we get now in a straightforward
way:
$$\eqalign{
&\sum_j\ dm^2_j{\partial\over\partial m^2_j}\ {}^{(0)}
P(T)=\sum_j\ dm^2_j{\partial\over\partial m^2_j} {}^{(0)}P(T)^{int}\cr
&-{C\over 2}dm^2_2\int{d^nk\over(2\pi)^n}n(k)2\pi\delta
(k^2-m^2_2),\cr}\eqno(A.23)$$
where ${}^{(0)}P(T)^{int}$ is the interaction pressure as in (2.17)
and the last term on the right-hand side is $dm^2_2$
times the derivative with respect to $m^2_2$ of the pressure of the free
photons of renormalised mass $m_2$. In deriving (A.23) we have
made use of the relations
$$\eqalign{
&{\rm Im}\ \pi''(m^2_2)=0,\cr
&{\rm Im}\ {\partial\over\partial m^2_1}\pi'(m^2_2)=0\cr}\eqno(A.24)$$
which are valid for stable photons and electrons, that is for
$$0\leq m^2_2< 4m^2_1\eqno(A.25)$$

%\topinsert
%\centerline{{\epsfxsize=100mm\epsfbox{8.ps}}}
%\vskip 1pt
%\centerline{Figure 8: The $m^2_1-m^2_2$-plane. The physical
%region corresponding to stable electrons and photons is
%$0\leq m^2_2\leq 4m^2_1$. $C_1$ is an example
%of an the integration path
%chosen in (A.26).}
%\endinsert
\bigskip
\goodbreak
\medskip\immediate\closeout\rfile\writestoppt
\baselineskip=14pt{{\bf References}}\bigskip{\frenchspacing%
\parindent=20pt\escapechar=` \input refs.tmp\bigskip}\nonfrenchspacing
\bye